\documentclass[11pt,leqno]{article}

\usepackage{latexsym}
\usepackage{amsfonts}
\usepackage{amsmath}
\usepackage{amssymb}
\usepackage{pifont}

\newlength{\filength}
\newsavebox{\gcbox}
\sbox{\gcbox}{\framebox[\filength]{\rule{0ex}{2ex}}}

  \newtheorem{theorem}{Theorem}[section]
  \newtheorem{corollary}[theorem]{Corollary}
  \newtheorem{claim}[theorem]{Claim}

\newcommand\qedblob{\ding{113}}
\def\literalqed{{\ \nolinebreak\hfill\mbox{\qedblob\quad}}}

  \newtheorem{lemma}[theorem]{Lemma}

  \newtheorem{proposition}[theorem]{Proposition}
  \newtheorem{definition}[theorem]{Definition}

\newcount\hour  \newcount\minutes  \hour=\time  \divide\hour by 60
\minutes=\hour  \multiply\minutes by -60  \advance\minutes by \time
\def\mmmddyyyy{\ifcase\month\or Jan\or Feb\or Mar\or Apr\or May\or Jun\or Jul\or
  Aug\or Sep\or Oct\or Nov\or Dec\fi \space\number\day, \number\year}
\def\hhmm{\ifnum\hour<10 0\fi\number\hour :%
  \ifnum\minutes<10 0\fi\number\minutes}

\makeatletter
\def\@citex[#1]#2{\if@filesw\immediate\write\@auxout{\string\citation{#2}}\fi
  \def\@citea{}\@cite{\@for\@citeb:=#2\do
    {\@citea\def\@citea{,\linebreak[0]}\@ifundefined
       {b@\@citeb}{{\bf ?}\@warning
       {Citation `\@citeb' on page \thepage \space undefined}}%
\hbox{\csname b@\@citeb\endcsname}}}{#1}}
\newcommand{\singlespacing}{\let\CS=
\@currsize\renewcommand{\baselinestretch}{1}\tiny\CS}
\newcommand{\singlespacingplus}{\let\CS=
\@currsize\renewcommand{\baselinestretch}{1.25}\tiny\CS}
\newcommand{\doublespacing}{\let\CS=
\@currsize\renewcommand{\baselinestretch}{1.75}\tiny\CS}
\newcommand{\extradoublespacing}{\let\CS=
\@currsize\renewcommand{\baselinestretch}{1.9}\tiny\CS}
\newcommand{\draftspacing}{\let\CS=
\@currsize\renewcommand{\baselinestretch}{2.0}\tiny\CS}
\newcommand{\hugedraftspacing}{\let\CS=
\@currsize\renewcommand{\baselinestretch}{2.4}\tiny\CS}
\newcommand{\normalspacing}{\singlespacing}
\makeatother%

\normalspacing

\newcommand{\naturalnumber}{\ensuremath{{  \mathbb{N} }}}



\newcommand{\p}{{\rm P}}

\newcommand{\np}{{\rm NP}}

\newcommand{\conp}{{\rm coNP}}

\newcommand{\thetatwo}{\ensuremath{\Theta_2^p}}

\newenvironment{proofs}{\noindent{\bf Proof.}\hspace*{1em}}{\literalqed\bigskip}

\newcommand{\sproof}{\noindent{\bf Proof.}\hspace*{1em}}

\newcommand{\sproofof}[1]{\noindent{\bf Proof of {#1}.}\hspace*{1em}}
\newcommand{\eproofof}[1]{\noindent{\hspace*{0.1in} \hfil \hfill \mbox{\literalqed{} {#1}}}\quad\bigskip}

\newcommand{\pair}[1]{\mathopen({#1}\mathclose)}
\newcommand{\manyone}{\ensuremath{\,\leq_m^p\,}}
\newcommand{\manyonetext}{\ensuremath{\mbox{$\leq_m^p$}}}

\newcommand{\sigmastar}{\ensuremath{\Sigma^\ast}}

\newcommand{\condition}{\,\mid \:}

\def\nats{\naturalnumber}

\def\parpair#1{{{(\!\!~#1~\!\!)}}}

\newcommand{\hybrid}{{{\mathit hybrid}}}

\newcommand{\plurality}{\mbox{\rm{}plurality}}
\newcommand{\condorcet}{\mbox{\rm{}Condorcet}}
\newcommand{\election}{{{\mathcal{E}}}}
\newcommand{\electiondefault}{{{\mathcal{E}_{\mbox{\protect\scriptsize\rm default}}}}}

\newcommand{\electionnotallone}{{{\mathcal{E}_{\mbox{\protect\scriptsize\rm not-all-one}}}}}

\newcommand{\electionnull}{{{\mathcal{E}_{\mbox{\protect\scriptsize\rm null}}}}}

\newcommand{\electiononefirst}{{{\mathcal{E}_{\mbox{\protect\scriptsize\rm first}}}}}
\newcommand{\electionlast}{{{\mathcal{E}_{\mbox{\protect\scriptsize\rm last}}}}}

\newenvironment{algorithmus}{\begin{list}
   {{\bf Step~\arabic{alg}:}}
   {\usecounter{alg}}}{\end{list}}
\newcounter{alg}

\newenvironment{case}{\begin{list}
   {{\bf Case~\arabic{casenumber}:}}
   {\usecounter{casenumber}}}{\end{list}}
\newcounter{casenumber}

\title{Hybrid Elections Broaden Complexity-Theoretic Resistance to
Control\thanks{
A preliminary version was presented at the COMSOC-06 workshop
and a preliminary, seven-page version
of this paper appeared in IJCAI-07~\cite{hem-hem-rot:c:hybrid}.}}

\author{Edith Hemaspaandra\thanks{URL:
\mbox{\tt{}www.cs.rit.edu/\mbox{\tiny$\sim\,$}eh}.
Supported in part by grants NSF-CCR-0311021
and 
NSF-IIS-0713061,
by the
Alexander von Humboldt Foundation's
TransCoop program, and by
a Friedrich Wilhelm Bessel Research Award.
Work
done in part while visiting
Heinrich-Heine-Universit\"{a}t D\"{u}sseldorf.
} \\
Department of Computer Science \\
Rochester Institute of Technology \\
Rochester, NY 14623, USA \and 
Lane A. Hemaspaandra\thanks{URL:
\mbox{\tt{}www.cs.rochester.edu/u/lane}.
Supported in part by grant
NSF-CCF-0426761, 
by the
Alexander von Humboldt Foundation's
TransCoop program, and by
a 
Friedrich Wilhelm Bessel Research Award.
Work
done in part while visiting
Heinrich-Heine-Universit\"{a}t D\"{u}sseldorf.
}
\\Department of Computer Science \\
University of Rochester \\
Rochester, NY 14627, USA
\and
J\"{o}rg Rothe\thanks{URL: 
\mbox{\tt{}ccc.cs.uni-duesseldorf.de/\mbox{\tiny$\sim\,$}rothe}.
Supported in part by grants 
DFG-RO-1202/\{\mbox{9-1},\mbox{9-3},\mbox{11-1},\mbox{12-1}\}, 
the European Science Foundation's EUROCORES program LogICCC, and by the
Alexander von Humboldt Foundation's
TransCoop program.} \\
Institut f\"{u}r Informatik \\
Heinrich-Heine-Universit\"{a}t D\"{u}sseldorf \\
40225 D\"{u}sseldorf, Germany
}

\date{September 26, 2008
}

\begin{document}

\sloppy

\maketitle

\begin{abstract}
Electoral control refers to attempts by an election's
organizer (``the chair'') to influence the outcome by
adding/deleting/partitioning voters or candidates.  The groundbreaking
work of Bartholdi, Tovey, and Trick~\cite{bar-tov-tri:j:control} on
(constructive) control proposes computational complexity as a means of resisting
control attempts: Look for election systems where the chair's task in
seeking control is itself computationally infeasible.

We introduce and study a method of combining two or more
candidate-anonymous election schemes in such a way that the combined
scheme possesses all the resistances to control (i.e., all the
$\np$-hardnesses of control) possessed by \emph{any} of its
constituents: It combines their strengths.  From this and new
resistance constructions, we prove for the first time that there
exists an election scheme that is resistant to all twenty standard
types of electoral control.
\end{abstract}

\section{Introduction}
\label{s:intro}

Elections are a way of, from a collection of voters' (or agents')
individual preferences over candidates (or alternatives), selecting a
winner (or 
outcome).  The importance of and study of elections is
obviously central in political science, but also spans such fields as
economics, mathematics, operations research, and computer science.
Within computer science, the applications of elections are most
prominent in distributed AI, most particularly in the study of
multiagent systems.  For example, voting has been concretely proposed
as a computational mechanism for
planning~\cite{eph-ros:c:clarke-tax,eph-ros:c:multiagent-planning} and
has also been suggested as an approach to collaborative
filtering~\cite{gil-hor-pen:c:collaborative-filtering}.  However,
voting also has received attention within the study of systems. After
all, many distributed algorithms must start by selecting a
leader,
and election techniques have also been proposed to attack the web page
rank aggregation problem and the related issue of lessening the spam
level of results from web
searches~\cite{dwo-kum-nao-siv:c:rank-aggregation,fag-kum-siv:c:similarity-search}.
Indeed, in these days of a massive internet with many pages, many
surfers, and many robots, of intracorporate decision-making
potentially involving electronic input from many
units/individuals/warehouses/trucks/sources, and more generally of
massive computational settings including many actors, it is easy to
note any number of situations in which elections are natural and in
which the number of candidates and/or voters might be massive.  For
example, suppose amazon.com were to select a ``page of the week'' via
an election where the candidates were all its web pages and the voters
were all visiting surfers (with preferences inferred from their
page-viewing times or patterns); such an election would have an
enormous number of candidates and voters.  All these applications are
exciting, but immediately bring to a theoretician's mind the worry of
whether the complexity of implementing election systems is
satisfyingly low and whether the complexity of distorting (controlling
or manipulating) election systems is reassuringly high.

Along these lines, though in a time when the importance of large-scale
elections was less crystal clear than it is today, a remarkable series
of papers by Bartholdi, Orlin, Tovey, and
Trick~\cite{bar-tov-tri:j:who-won,bar-tov-tri:j:manipulating,bar-orl:j:polsci:strategic-voting,bar-tov-tri:j:control}
pioneered the study of the complexity of the winner, control, and
manipulation problems for election systems
(see~\cite{fal-hem-hem-rot:btoappearWithTrPtr:richer} 
for a related survey).  Of these, the winner and
manipulation problems have since been studied particularly
intensely.  It is known, for example, that even for some quite
natural, simple-to-state election systems---such as those of Lewis
Carroll~\cite{dod:unpub:dodgson-voting-system},
Kemeny (see~\cite{kem-sne:b:polsci:mathematical-models}), and
Young~\cite{you:j:extending-condorcet,lev-you:j:condorcet}---checking whether a given
candidate wins is complete for the ``parallel
access to
$\np$'' level of 
the polynomial
hierarchy
\cite{hem-hem-rot:j:dodgson,rot-spa-vog:j:young,hem-spa-vog:j:kemeny}.

Since the complexity of elections is a topic whose importance has made itself
clear, it is natural to ask whether the standard tools and techniques of
complexity-theoretic study exist in the context of elections.  One important
technique in complexity is the combination of problems.  For example, for sets
in complexity theory, a standard approach to combination is the join (also
known as the disjoint union and as the marked union): $A \oplus B = \{0x
\condition x \in A\} \cup \{1y \condition y \in B\}$.  Even this simple
example is not as innocent as it seems.  Though the join indeed has many nice
properties regarding respecting simplicity (e.g., if $A \in \p$ and $B \in
\p$, then $A \oplus B \in \p$) and preserving hardness (e.g., if $C \manyone
A$ or $C \manyone B$, then $C \manyone A \oplus B$), it is known that the join
has the quite unexpected ability of lower complexity within the extended low
hierarchy (\cite{hem-jia-rot-wat:j:join-lowers}; 
that result should probably best be
taken as suggesting that the definition of the extended low hierarchy---back
in 1986~\cite{bal-boo-sch:j:low}---was itself somewhat unnatural, or at least
left itself open to this unusual behavior).

In some sense, our work in this paper can be thought of as simply
providing, for elections, an analog of the join.  That is, we will
propose a method
of combining two (or more) elections in a way
that will maintain desirable simplicity properties (e.g., if all of
the constituent elections have polynomial-time winner algorithms then
so will our combined election) while also inheriting quite
aggressively desirable hardness properties (we will show that any
resistance-to-control---in the sense that is
standard~\cite{bar-tov-tri:j:control} and that we will provide a
definition of later---possessed by even one of the constituent
elections will be possessed by the combined election).  One cannot
directly use a join to achieve this, because the join of two sets
modeling elections is not itself an election.  Rather, we must find a
way of embedding into election specifications---lists of voter
preferences over candidates---triggers that both allow us to embed and
switch between all the underlying election systems and to not have
such switching go uncontrollably haywire when faced with electoral
distortions such as adding/deleting/partitioning voters/candidates,
since we wish hardness with respect to control by such mechanisms to
be preserved.

We above have phrased this paper's theme as the development of a way
of combining multiple election systems---and in doing so, have
desirable types of simplicity/complexity inheritance.  However, this
paper also has in mind a very specific application---both for its own
interest and as a sounding board against which our election
hybridization scheme can be tested.  This application is the control
of election systems.

In election control, we ask whether an election's organizer (the
chair) can by some specific type of manipulation of the election's
structure (adding/deleting/partitioning voters/candidates) cause a
specified candidate to be the (unique) winner.  As mentioned earlier,
the complexity-theoretic study of control was proposed by Bartholdi,
Tovey, and Trick in 1992~\cite{bar-tov-tri:j:control}.  We will closely follow
their model.  In this model, the chair is assumed to have knowledge of
the vote that will be cast by each
voter,\footnote{\cite{bar-tov-tri:j:control}, in this regard, takes
the view that all voters vote sincerely, and makes the good point
regarding the chair's global knowledge that such global knowledge is
perhaps not a realistic assumption but rather is a ``conservative''
one---that they show that even in the face of an omniscient chair, the
control problems for certain voting systems are intractable.  We
heartily agree with the latter point, but regarding the former point
mention that one does not need to assume voter sincerity.  Rather,
their model simply assumes that the chair knows how each voter will
vote, and it is irrelevant whether such votes are or are not
sincere---at least if our model is that the votes are fixed before
the chair makes his or her decisions regarding how to assert control.}
and there are ten different types of control (candidate addition,
candidate deletion, voter addition, voter deletion, partition of
candidates, run-off partition of candidates, and partition of
voters~\cite{bar-tov-tri:j:control}---and for each of the three
partition cases one can have subelection ties promote or can have
subelection ties eliminate,
see~\cite{hem-hem-rot:j:destructive-control}).

Though Bartholdi, Tovey, and Trick's control paper is lovely and
powerful, very few subsequent papers have extended beyond their
results.  One particular one that has is a paper of Hemaspaandra,
Hemaspaandra, and Rothe~\cite{hem-hem-rot:j:destructive-control} that
asks the same questions for \emph{destructive control}, i.e., when the
chair's goal is to preclude a given candidate from being the
(unique)
winner.  (One might ask if such study is a waste of time, given that
any sane chair would vastly prefer to choose the winner than to just
keep some particular hated choice from being the winner.  However,
\cite{hem-hem-rot:j:destructive-control} shows that for some standard
election systems and types of control the former problem is
intractable yet the latter is in~$\p$.  That is, spoiling a
candidate's chances can be easier than outright controlling who wins.)

Of course, the dream case would be to find an election system that has
the desirable property of having a polynomial-time algorithm for
evaluating who won, but that also has the property that for every
single one of the twenty standard types of control 
(namely, the standard ten types, each for the constructive
and destructive cases)
it is computationally
infeasible ($\np$-hard) to assert such control.  Unfortunately, no
system yet has been proven resistant to all twenty 
types of control.  In
fact, given that broad ``impossibility'' results exist for niceness of
preference aggregation systems 
(e.g., Arrow's Theorem~\cite{arr:b:polsci:social-choice})
and for nonmanipulability of election systems (e.g., the
Gibbard--Satterthwaite and Duggan--Schwartz Theorems
(\cite{gib:j:polsci:manipulation,sat:j:polsci:manipulation,dug-sch:j:polsci:gibbard}, 
see also~\cite{tay:b:polsci:social-choice-manipulation})),
one might even momentarily wonder whether the ``dream case'' mentioned
above can be proven impossible via proving a theorem of the following
form:
\begin{quote}
``For no election system whose winner complexity is in P             
  are all twenty types of control $\np$-hard.''
\end{quote}
However, such a claim is proven impossible by our work: Our hybrid
system in fact will allow us to combine \emph{all} the resistance
types of the underlying elections.  And while doing so, it will
preserve the winner-evaluation simplicity of the underlying elections.
Thus, in particular, we conclude that the ``dream case'' holds: There
are election systems---for example, our hybridization of plurality,
Condorcet, and an election system we call $\electionnotallone$
(which will be defined in Section~\ref{sec:resistance})---that 
are resistant to all twenty types of
control (Theorem~\ref{t:all-twenty}).

This paper (in its conference version) was the first to prove any
election system to be resistant to even the ten \emph{constructive}
control types.  And it was only in 2008, in the work
of~\cite{fal-hem-hem-rot:c:copeland-alpha}, that it was shown that
some natural system also had those ten constructive-control
resistances.  (The largest resistance count among the 
twenty standard control types that has yet been obtained by any
work other than that of the present paper---which obtains all twenty
resistances---was obtained by Erd\'{e}lyi, Nowak, and
Rothe~\cite{erd-now-rot:t-With-MFCS08-Ptr:sp-av}, who
proved that a variant of approval voting has seventeen of the twenty
resistances.  However, to do so that paper adopts a different
model---one in which each voter's preference consists of \emph{both} a
strict ordering and an approval vector.%
)  It remains an open issue
whether any previously studied (i.e., ``natural'') 
system has all twenty
resistances.

Our hybridization system takes multiple elections and maintains their
simplicity while inheriting each resistance-to-control possessed by
any one of its constituents.  Thus, it in effect unions together all
their resistances---thus the ``broaden'' of our title.\footnote{We mention 
that in the quite different setting of election manipulation
(which regards not actions by the chair but rather which regards
voters altering their preferences in an attempt to influence who
becomes the winner)~\cite{bar-tov-tri:j:manipulating}, there has been
some work by Conitzer and Sandholm~\cite{con-san:c:voting-tweaks}
regarding making manipulation hard, even for systems where it is not
hard, by changing the system by going to a two-stage election in which
a single elimination pre-round is added, and Elkind and
Lipmaa~\cite{elk-lip:c:hybrid-voting-hardness-manipulation} have
generalized this to a sequence of elimination rounds conducted under
some system(s) followed by an election under some other system.
Though the latter paper like this paper uses the term ``hybrid,'' the
domains differ sharply and the methods of election combination are
nearly opposite: Our approach (in order to broaden resistance to
control) embeds the election systems in parallel and theirs (in order
to fight manipulation) strings them out in sequence.  
}
The 
work most closely related to that of this paper is the
constructive control-defining work of Bartholdi, Tovey, and
Trick~\cite{bar-tov-tri:j:control},
the destructive control-defining work of
Hemaspaandra, Hemaspaandra, and
Rothe~\cite{hem-hem-rot:j:destructive-control},
and the control studies of 
Erd\'{e}lyi, Nowak, and 
Rothe~\cite{erd-now-rot:c:sp-av,erd-now-rot:t-With-MFCS08-Ptr:sp-av},
Faliszewski et 
al.~\cite{fal-hem-hem-rot:c:llull,fal-hem-hem-rot:c:copeland-alpha},
and Procaccia, Rosenschein, and Zohar~\cite{pro-ros-zoh:c:multiwinner}.
Work on bribery is 
related to this paper, in the 
sense that bribery can be viewed 
as sharing aspects of both manipulation and 
control~\cite{fal-hem-hem:c:bribery,fal:c:nonuniform-bribery}.
Of course, all the
classical~\cite{bar-tov-tri:j:who-won,bar-tov-tri:j:manipulating,bar-orl:j:polsci:strategic-voting}
and recent papers (of which we particularly point out, for its broad
framework and generality, the work of Spakowski and
Vogel~\cite{spa-vog:c:theta-two-classic}) on the complexity of election
problems share this paper's goal of better understanding the
relationship between complexity and elections.

This paper is organized as follows. Section~\ref{s:prelims} provides
the needed definitions regarding elections and control, defines and
discusses our hybridization system, and defines and discusses the
notions of inheritance that we will use to measure whether or not
hybridization maintains or destroys desirable properties.
Section~\ref{s:sriv} discusses the behavior of our hybrid system with
respect to inheriting resistance ($\np$-hardness) to control, and also
discusses the related notion of inheritance of susceptibility.  Then
it discusses the inheritances and noninheritances of our hybrid system
with respect to vulnerability (having $\p$-time control algorithms),
and also discusses the related notion of immunity.  
Appendix~\ref{a:defer} presents some proofs
deferred from Section~\ref{ss:vulnerable}.

\section{Elections, Control, Hybridization, and Inheritance:
Definitions and Discussion}
\label{s:prelims}

\subsection{Elections}

An election system (or election rule
or election scheme or voting system) $\election$ is simply a mapping from (finite
though arbitrary-sized) sets (actually, mathematically, they are
multisets)
$V$ of votes (each a preference order---strict, transitive, and
complete---over a finite candidate
set\footnote{\label{f:88-candidate-argument}A subtle point is that, to be
  very careful, we should mention that though we above speak of $\election$ as
  if its input was just the votes (over a candidate set), to be utterly formal
  we tacitly view the candidate set itself as being also an input.  It might
  be natural to think that taking such a view is unneeded, since we can infer
  the candidate set from the votes.  However, that reasoning fails---though
  this boundary case seems to have been overlooked in some earlier work---if
  the cardinality of the vote set is zero.  For example, the election rule
  under which all candidates always win very fundamentally needs to, when
  there are zero voters, know what the candidate set is, so it can know what
  winner names to output.  Nonetheless, we will be careful in defining our
  hybrid scheme not to take unfair advantage of the fact that $C$ is part of
  our input, formally speaking.

In the context of subelections we will sometimes speak of elections
$\parpair{C,V}$ where $C$ is a strict subset of the candidates that
occur in $V$.  In such cases, we view the preference orders
in $V$ as being restricted to $C$.})
to (possibly empty, possibly nonstrict) subsets of the
candidates.\footnote{\label{f:66-social-choice}As is typical in papers on the
  complexity of voting, we study the case not of elections (actually,
  full-blown social choice functions) that output a full preference order
  (except with ties allowed), but rather we study the bare-bones but in
  practice very central case of elections where the focus is purely on
  selecting the winners.
  
  As a somewhat arcane side comment, we mention that though many earlier papers
  imply that the latter
  is a special case of the former, this is not strictly true unless one
  restricts oneself to election rules always having at least one winner.
  Without that restriction, the problem is that the trick of declaring the top
  equality group of an output ``preference order except with ties allowed'' to
  be the winners breaks down since it has no way to declare that there are no
  winners (note that having all be equal would make them all winners; of
  course, we could when there are three or more candidates use, for example,
  $c>b>a$ to code both $c>b>a$ and $c>a>b$, since they both have the same
  winner set, ``$\{c\}$,''
  and we would thus free up $c>a>b$ to denote ``special
  case: no one wins''\ldots\ but this is such an exceedingly unnatural
  approach that we would not want to say that it shows that winner/loser
  elections with three or more candidates are a ``special case'' of general
  social choice functions).}
All votes in a given $V$ are over the same candidate set, but different
$V$'s of course can be over different (finite) candidate sets.
Each candidate that for a given set of votes is in $\election$'s 
output is said to be a \emph{winner}.
If for a given input $\election$ outputs a set of cardinality one,
that candidate is said to be the \emph{unique winner}.  Election control
focuses on making candidates be unique winners and on precluding
them from being unique winners.

Throughout this paper, a voter's preference order will be exactly that: a
tie-free linear order over the candidates.  And we will discuss and hybridize
only election systems based on preference orders.\footnote{Thus we do not
  directly discuss systems such as approval voting that take very different
  inputs: lists tagging each candidate as approved or not approved.  However,
  we mention in passing that our hybridization approach works perfectly well
  for such systems and for them yields---if one codes the input as candidate
  lists tagged with an approval/disapproval bit---the same main results: The
  hybrid inherits all the resistance types of its constituent systems and yet
  has a polynomial-time winner algorithm if they all do.  Thus the only reason
  we throughout this paper fix our systems as having preference orders as
  inputs is that it wouldn't even be particularly meaningful to try to
  hybridize
  together systems with conflicting input types, and by focusing just on
  preference-order-based preferences we prevent that from ever happening.}
Following a convention that dates at least back to the early 1980s, 
we often will refer to the elements of $V$ as ``voters'' rather
than ``votes.''

We now define two common election systems, plurality voting and Condorcet
voting. In \emph{plurality voting}, the winners are the candidates
who are ranked first the most.
In \emph{Condorcet voting}, the winners are all
candidates (note: there can be at most one and there might be zero)
who strictly beat each other candidate in head-on-head
majority-rule
elections (i.e., get \emph{strictly} more than half the
votes in each such election).
For widely used systems such as plurality voting, we
will write plurality rather than $\election_{\mbox{\protect\scriptsize\rm
    plurality}}$.

We say that an election system $\election$ is \emph{candidate-anonymous} if
for every pair of sets of votes $V$ and $V'$,
$\|V\| = \|V'\|$, such that $V'$ can be
created from $V$ by applying some one-to-one mapping $h$ from the candidate names
in $V$ onto new candidate names in $V'$ (e.g., each instance of ``George'' in
$V$ is mapped by $h$ to ``John'' in $V'$ and each instance of ``John'' in $V$
is mapped by $h$ to ``Hillary'' in $V'$ and each instance of ``Ralph'' in $V$
is mapped by $h$ to ``Ralph'' in~$V'$) it holds that
\[\election(V') = \{c' \condition (\exists c \in \election(V))\, [h(c) = c']\}.
\]
Informally put, candidate-anonymity says that the strings we may use to name
the candidates are all created equal.  Note that most natural systems
are candidate-anonymous.  For example, both
the election systems mentioned immediately above---plurality-rule 
elections and the election system of Condorcet---are 
candidate-anonymous.

We mention that
everywhere in this paper where we use assumptions of
candidate-anonymity (namely, in most of our inheritance results), our results
would even hold under the far weaker assumption of
candidate-affine-invariance, defined as 
follows.\footnote{One should 
be a bit careful here.  In many such cases, our existing proof
already handles the case of 
candidate-affine-invariance, since the existing proof
uses a transform of the form used in 
defining (as we are about to do in the main text) 
candidate-affine-invariance:
$h(c) = k_1 c + k_2$, $k_1 \in \nats^+$, $k_2 \in \nats$.  
However, in some cases, rethinking the result in terms of 
candidate-affine-invariance requires revising the proof, most typically to 
either explicitly invoke the type of mapping used in defining 
candidate-affine-invariance, or to change the words 
``candidate-anonymous''/``candidate-anonymity''
to the words 
``candidate-affine-invariant''/``candidate-affine-invariance'' 
in some places, or to 
keep in mind that when considering this change one also should 
view all the underlying definitions---such as inheritance---as being
framed in terms of candidate-affine-invariance (though, of course,
in any case candidate-anonymity implies candidate-affine-invariance, so
the remapping of the definition of inheritance is not truly 
needed when the inheritance appears only in the hypothesis of 
a particular theorem).  
As a specific example,
note that to make 
Theorem~\ref{t:special}'s proof work for the case of 
candidate-affine-invariance,
one would have to create that proof's set $D$ from that proof's set
$C$ not by just using any mapping that ensures that $D$ and $C$ are disjoint,
but rather by choosing a mapping that achieves disjointness
and that also has the form
$h(c) = k_1 c + k_2$, $k_1 \in \nats^+$, $k_2 \in \nats$;  this can
easily be done, e.g., by taking $k_1 = 1$ and $k_2$ sufficiently 
large.}
We say that an election
system $\election$ is \emph{candidate-affine-invariant} if for
every pair of sets of votes $V$ and $V'$ such that $V'$ can be created from
$V$ by applying some mapping $h$ of the form 
$h(c) = k_1 c +
k_2$, $k_1 \in \nats^+$, $k_2 \in \nats$, from the candidate names in $V$ to
the new candidate names in~$V'$, it holds that
$\election(V') = \{c' \condition (\exists c \in \election(V))\, [h(c) = c']\}.$
Note that when we treat candidate names as natural numbers (both above, and
also later in defining our hybrid scheme), we are implicitly using the
standard bijection between strings from $\sigmastar$ (potential candidate
names) and natural numbers, namely the $i$th string in lexicographical order
is associated with the natural number $i-1$.  Though as just mentioned
candidate-affine-invariance suffices for all our results,
we nonetheless use
candidate-anonymity in our definitions and statements as candidate-anonymity
seems a more natural notion. To avoid any confusion,  we stress that
$k_1$ is limited to $\nats^+$ and $k_2$ is limited to $\nats$,
and so we are using the term ``affine'' somewhat improperly.

Finally, we mention that though related in flavor, candidate-anonymity is not
identical to the political science notion of (candidate-)neutrality
(see~\cite{arr:b:polsci:social-choice}),
which is defined by saying that permuting
the names of an election's candidates does not affect the
outcome.\footnote{Arrow~\cite{arr:b:polsci:social-choice} cites the 
origin of this notion to a paper by
Guilbaud~\cite{gui:j:theories-of-aggregation}
and defines the Condition of Neutrality as follows:
``Let $T(x)$ be a one-one transformation of the set of alternatives into
itself which preserves all individual orderings.  Let the environment $S$
be transformed into $S'$ by the transformation~$T$.  Then the social 
choice from $S$, $C(S)$, is transformed by $T$ into the social choice,
$C(S')$, from the environment~$S'$.''
\label{f:neutral}}
Candidate-anonymity implies neutrality but the converse does not in general
hold as shown by the following example.  The election system that ignores all
the votes and declares all candidates winners if all candidates' names viewed
as strings are of the same length and otherwise declares no candidates winners
is neutral but not candidate-anonymous (``Bob'' and ``Joy'' both win in the
two-candidate election with (irrelevant) vote set $V$ but changing ``Bob'' to
``Bobby'' changes the outcome).  Candidate-anonymity, with its view of names
as being
coded as arbitrary strings, is a more ``computer science''-y
notion---focusing on actual codings, since sets are defined via (and machines
accept) actual coded input strings.

\subsection{Our Hybridization Scheme: Definition and Discussion}
\label{ss:scheme}

We now define our basic hybridization scheme, $\hybrid$.
To avoid long
discussions about what properties of what we will call the ``default''
election rule are not needed or not inherited, we will first define a more
general scheme, $\hybrid$-${\mathit base}$,
and then will in defining $\hybrid$
simply fix the ``default'' election rule to be the same as the last regular
constituent election rule.  By doing so, our results will be stated in
slightly less general ways than what actually holds, but the theorem
statements will be simpler and clearer.

\begin{definition}
  Let $\election_0, \election_1, \ldots , \election_{k-1}$, and
  $\electiondefault$ be election rules that take as input voters' preference
  orders.  We define
  $\hybrid$-${\mathit base}(\election_0, \election_1, \ldots ,
  \election_{k-1},\electiondefault)$ to be the election scheme
  that does the following:
\begin{quote}
  If there is at least one candidate and all candidate names\footnote{Viewed
    as natural numbers via the standard bijection between $\sigmastar$ and
    $\nats$ mentioned in the discussion of candidate-anonymity.} are
  congruent, modulo~$k$, to $i$ (for some~$i$, $0 \leq i \leq k-1$) then use
    election rule~$\election_i$.  Otherwise use $\electiondefault$.
\end{quote}
We define $\hybrid(\election_0, \election_1, \ldots , \election_{k-1})$ to be
the system $\hybrid$-${\mathit base}(\election_0, \election_1, \ldots ,
\election_{k-1}, \election_{k-1})$.  We say that $\election_0, \election_1, \ldots ,
\election_{k-1}$ are the \emph{constituents} of $\hybrid(\election_0,
\election_1, \ldots , \election_{k-1})$.
\end{definition}

Having defined our system there is much to discuss.  Why did we choose this
system?  What are its properties?
What other approaches did we choose not to use, and why?
What aspects of the input is our method for
switching between election systems using, and what aspects is it choosing not
to exploit, and what are the costs associated with our choices?

As to the properties of this system, Section~\ref{s:sriv} is devoted to that,
but most crucially we will see that this system possesses every
resistance-to-control property possessed by even one of its constituents.  And
this will hold essentially due to the fact that $\hybrid$ is a close analog of
the effect of a join: It splices the constituents together in such a way that
key questions about the constituent systems can easily be many-one
polynomial-time reduced ($\manyonetext$-reduced
or reduced, for short)
to questions about their hybrid.\footnote{We mention
  in passing that this is one of the reasons we focus on $\hybrid$ rather than
  $\hybrid$-${\mathit base}$.  It is possible to
  construct artificial systems for
  which even the winner problem for a system $\election_{D}$ does not
  $\manyonetext$-reduce to $\hybrid$-${\mathit base}(\election_0, \election_1,
  \ldots , \election_{k-1}, \election_{D})$.
}

As to why we chose this particular system, note that $\hybrid$ ``switches''
between constituent systems via wildly redundant information.  This will let
us keep deletions/partitions 
of voters/candidates from causing a switch between
the underlying systems (if the starting state routed us to a nondefault case).
Note that some other approaches that one might take are more sensitive to
deletions.  For example, suppose we wanted to hybridize just two election
systems and decided to do so by using the first election system exactly if the
first voter's most disliked candidate's name is lexicographically less than
the first voter's second-most-disliked candidate's name.
Note that if, as part of our control problem, that voter is deleted, that might
suddenly change the system to which the problem is routed.  Or, as another
example, if we use the modulo~$k$ value of the name of the lexicographically
smallest candidate to control switching between the $k$ election systems, then
that hybridization approach would be very sensitive to jumping between systems
if, as part of our control problem, that candidate is deleted.  These examples
give some idea of why we chose the approach we did, though admittedly even it
can in some cases be nudged into jumping between systems---but at least this
happens in very limited, very crisply delineated cases and in ways that we
will generally be able to appropriately handle.

Finally, we come to what we allow ourselves to use to control the switching,
what we choose not to use, and what price we pay for our choices.  What we use
(as is allowed in the~\cite{bar-tov-tri:j:control} model) are the candidates'
names and only the candidates' names.  We use absolutely nothing else to
control switching between elections.  We do not use voters' names.  Indeed, in
the~\cite{bar-tov-tri:j:control} model that we follow, voters (unlike
candidates) do not even have names.  But since the votes are input as a list,
their ordering itself could be used to pass bits of information---e.g., we
could look at whether the first vote in the list viewed as a string is
lexicographically less than the last vote in the list viewed as a string.  We
in no way ``cheat'' by exploiting such input-order information, either for the
votes or for the list of candidates (as per
Footnote~\ref{f:88-candidate-argument}, and as
per~\cite{bar-tov-tri:j:control}, formally the candidate set is passed in
separately to cover a certain boundary case).  Our ``switch'' is based
purely
on candidates' names and just candidates' names.  This also points to the
price we pay for this choice: Even when all its constituent elections are
candidate-anonymous, $\hybrid$ may not possess candidate-anonymity.

\subsection{Types of Constructive and Destructive Control}

Constructive
control problems ask whether a certain class of actions by the election's
chair can make a specified candidate the election's unique winner.
Constructive control was first defined and studied by Bartholdi, Tovey, and
Trick~\cite{bar-tov-tri:j:control}.  Destructive control problems ask whether
a certain class of actions by the election's chair can make a specified
candidate fail to be a unique winner of the election.  Destructive control was
defined and studied by Hemaspaandra, Hemaspaandra, and
Rothe
\cite{hem-hem-rot:cOutByJournalToAppear:destructive-control,hem-hem-rot:j:destructive-control}, 
and in the different context
of electoral manipulation destruction was introduced even earlier by Conitzer,
Lang, and
Sandholm~\cite{con-san:c:few-candidates,con-lan-san:c:how-many-manipulate,con-lan-san:j:when-hard-to-manipulate}.

Bartholdi, Tovey, and Trick's \cite{bar-tov-tri:j:control}
groundbreaking paper defined seven types of
electoral control.
Among those seven, three are partition problems for which there are two
different natural approaches to handling ties in subelections
(see~\cite{hem-hem-rot:j:destructive-control} which introduced these
tie-handling models for this context): eliminating tied subelection winners
(the ``TE'' model) or promoting tied subelection winners (the ``TP'' model).
Thus, there are $(7-3) + 2\cdot 3 = 10$ different standard types of
constructive control, and there are essentially the same ten types of
destructive control.

Since it is exceedingly important to not use a slightly different problem
statement than earlier work whose results we will be drawing on, we will state
the seven standard 
constructive
control types (which become ten with the three partition
control types each having both ``TE'' and ``TP'' versions) 
and their destructive analogs
using word-for-word
definitions from
\cite{hem-hem-rot:j:destructive-control,hem-hem-rot:cOutByJournalToAppear:destructive-control,hem-hem-rot:t:destructive-control},
which themselves are based closely and often identically on
\cite{bar-tov-tri:j:control} (see the discussion
in~\cite{hem-hem-rot:j:destructive-control}).

Though $V$, the set of votes, is conceptually a multiset as in the previous
related work, we take the view that the votes are input as a list (``the
ballots''), and in particular are not directly input as a multiset in which
cardinalities are input
in binary (though we will mention later that our main result about $\hybrid$
holds also in that quite different model).

\subsubsection*{Control by Adding Candidates}

As is common, we state our decision problems as ``Given'' instances,
and a related Yes/No question.  The language in each case is the set
of all instances for which the answer is Yes.  (So
we are studying the complexity of recognizing the set of 
instances on which control is possible.)
Since in each control scenario, the
``Given'' instance is identical for the constructive and the destructive case,
we state it just once and then state the corresponding two questions, one
for constructive and one for destructive control.

\begin{description}
\item[Given:] A set $C$ of qualified candidates and a distinguished candidate
  $c \in C$, a set $D$ of possible spoiler candidates, and a 
  set $V$ of voters with preferences over $C \cup D$.
\item[Question (constructive):] Is there a choice of candidates from
  $D$ whose entry into the election would assure that $c$ is the
  unique winner?
\item[Question (destructive):] Is there a choice of candidates from
  $D$ whose entry into the election would assure that $c$ is not the
  unique winner?
\end{description}
This problem captures whether adding candidates can make a candidate the
victor or can block such.\footnote{The notion, as we have stated it,
is the notion proposed by 
Bartholdi, Tovey, and Trick~\cite{bar-tov-tri:j:control} as 
``Adding Candidates.''  However, 
\cite{fal-hem-hem-rot:c:llull,fal-hem-hem-rot:c:copeland-alpha}
recently defined and highly commended 
an eleventh control type---adding a number of candidates
for the ``limited'' case where a bound on the maximum number of candidates
that can be legally added is itself part of the input.  We mention that 
the techniques of the current paper apply perfectly well to that 
case too, and so in particular the techniques of the present paper
easily yield that there is a hybrid election system that is resistant
to all twenty-two control types (the twenty standard ones plus the 
constructive and destructive limited-candidate-addition cases).
The reason we can make this comment is that there are systems resistant
to constructive 
limited-candidate-addition 
and there are systems resistant to 
destructive
limited-candidate-addition;  for example, plurality has both 
these properties (see the discussion in the caption of 
Table~1 of~\cite{fal-hem-hem-rot:t3:llull}).}

\subsubsection*{Control by Deleting Candidates}

\begin{description}
\item[Given:] A set $C$ of candidates, a distinguished candidate $c \in C$, a
  set $V$ of voters, and a positive integer $k < \|C\|$.
\item[Question (constructive):] Is there a set of $k$ or fewer candidates in
  $C$ whose disqualification would assure that $c$ is the unique
  winner?
\item[Question (destructive):] Is there a set of $k$ or fewer candidates in
  $C - \{c\}$ whose disqualification would assure that $c$ is not
  the unique winner?
\end{description}
This problem models vote suppression.  (The choice to have a $k$ for candidate
deletion and voter addition/deletion but not for candidate addition is
somewhat arbitrary,
but was a decision made explicitly by Bartholdi, Tovey,
and Trick~\cite{bar-tov-tri:j:control}, and for consistency and comparison
purposes, we follow their choice.)
Note that deleting the distinguished candidate
$c$ in the destructive case, which would trivialize the problem, is not
allowed.

\subsubsection*{Control by Partition of Candidates}

\begin{description}
\item[Given:] A set $C$ of candidates, a distinguished candidate $c \in C$,
  and a set $V$ of voters.
\item[Question (constructive):] Is there a partition of $C$ into $C_1$
  and $C_2$ such that $c$ is the unique winner in the sequential
  two-stage election in which the winners in the subelection $\parpair{C_1,V}$
  who survive the tie-handling rule move forward to face the
  candidates in $C_2$ (with voter set~$V$)?
\item[Question (destructive):] Is there a partition of $C$ into $C_1$
  and $C_2$ such that $c$ is not the unique winner in the sequential
  two-stage election in which the winners in the subelection $\parpair{C_1,V}$
  who survive the tie-handling rule move forward to face the
  candidates in $C_2$ (with voter set $V$)?
\end{description}

\subsubsection*{Control by Run-off Partition of Candidates}

\begin{description}
\item[Given:] A set $C$ of candidates, a distinguished candidate $c \in C$,
  and a set $V$ of voters.
\item[Question (constructive):] Is there a partition of $C$ into $C_1$
  and $C_2$ such that $c$ is the unique winner of the election in
  which those candidates surviving (with respect to the tie-handling
  rule) subelections $\parpair{C_1,V}$ and $\parpair{C_2,V}$ have a run-off with voter
  set~$V$?
\item[Question (destructive):] Is there a partition of $C$ into $C_1$
  and $C_2$ such that $c$ is not the unique winner of the election in
  which those candidates surviving (with respect to the tie-handling
  rule) subelections $\parpair{C_1,V}$ and $\parpair{C_2,V}$ have a run-off with voter
  set $V$?
\end{description}
These two problems model two natural ways of conducting partitioned elections:
either via an elimination pre-round involving some candidates or as a run-off
structure (3-node tree).

\subsubsection*{Control by Adding Voters}

\begin{description}
\item[Given:] A set of candidates $C$ and a distinguished candidate $c \in C$,
  a set $V$ of registered voters, an additional set $W$ of yet
  unregistered voters (both $V$ and $W$ have preferences over~$C$), and a
  positive integer $k \leq \|W\|$.
\item[Question (constructive):] Is there a set of $k$ or fewer voters from $W$
  whose registration would assure that $c$ is the unique winner?
\item[Question (destructive):] Is there a set of $k$ or fewer voters from $W$
  whose registration would assure that $c$ is not the unique winner?
\end{description}

\subsubsection*{Control by Deleting Voters}

\begin{description}
\item[Given:] A set of candidates~$C$, a distinguished candidate $c \in C$, a
  set $V$ of voters, and a positive integer $k \leq \|V\|$.
\item[Question (constructive):] Is there a set of $k$ or fewer voters in $V$
  whose disenfranchisement would assure that $c$ is the unique winner?
\item[Question (destructive):] Is there a set of $k$ or fewer voters in $V$
  whose disenfranchisement would assure that $c$ is not the unique
  winner?
\end{description}

\subsubsection*{Control by Partition of Voters} 

\begin{description}
\item[Given:] A set of candidates~$C$, a distinguished candidate $c \in C$,
  and a set $V$ of voters.
\item[Question (constructive):] Is there a partition of $V$ into $V_1$ and
  $V_2$ such that $c$ is the unique winner in the hierarchical two-stage
  election in which the survivors of $\parpair{C,V_1}$ and $\parpair{C,V_2}$
  run against each other with voter set~$V$?
\item[Question (destructive):] Is there a partition of $V$ into $V_1$
  and $V_2$ such that $c$ is not the unique winner in the hierarchical
  two-stage election in which the survivors of $\parpair{C,V_1}$ and
  $\parpair{C,V_2}$ run against each other with voter set~$V$?
\end{description}
These three problems model control by addition, deletion, and partition of
voters.

\subsection{Immunity, Susceptibility, Vulnerability, and Resistance}
\label{s:prelims-isvr}

Again, to allow consistency with earlier papers and their results, we take
this definition from
\cite{hem-hem-rot:j:destructive-control,hem-hem-rot:cOutByJournalToAppear:destructive-control,hem-hem-rot:t:destructive-control},
with the important exception regarding resistance discussed below
Definition~\ref{d:four}.  It is worth noting that immunity and susceptibility
both are ``directional'' (can we change \emph{this}?)~but
that vulnerability and resistance are, in contrast,
outcome-oriented
(can we end up with \emph{this} happening?)~and complexity-focused.

\begin{definition}
\label{d:four}
  We say that a voting system is {\em immune to 
    control\/} in a given model of control (e.g., ``destructive 
control via adding candidates'') if
the model regards constructive control and 
it is never possible for the chair to 
by using his/her allowed model of control
change a
given candidate from being not a unique winner to being the
unique winner, or the model regards 
destructive control 
and it is never possible for the chair to 
by using his/her allowed model of control
change a
given candidate from being the unique winner to not 
being a 
unique winner.
If a system is not 
immune to a type of control, it is said to be 
\emph{susceptible} to that type of control. 

  A voting system 
  is said to be {\em (computationally)
    vulnerable to control\/} if it is susceptible to control 
  and the corresponding language problem is computationally
  easy (i.e., solvable in polynomial time).  

  A voting system 
is said to be {\em resistant to
    control\/} if 
it is susceptible to control 
but the corresponding language 
problem is computationally hard (i.e.,
  $\np$-hard).
\end{definition}

We have diverged from all previous papers by defining resistance as meaning
$\np$-hardness (i.e., $\np$-$\manyonetext$-hardness) rather than
$\np$-completeness (i.e., $\np$-$\manyonetext$-completeness).
Though revising existing notions always holds the danger of sowing confusion,
in this particular case the redefinition is compelling.
Under the \cite{bar-tov-tri:j:control} ``$\np$-complete'' definition, 
a problem whose control issue is $\np$-complete is resistant but a problem
whose control issue is NEEEE-complete---or even undecidable---would actually
be provably not resistant.  This is totally at odds with the intent
of the notion of being resistant to control.  
In~\cite{bar-tov-tri:j:control}, all their problems happened to be not just
$\np$-hard but also $\np$-complete, but that certainly is not a reason to
define a ``difficulty level'' concept in terms of a 
notion---$\np$-completeness---that encapsulates both a difficulty notion
($\np$-hardness) \emph{and a simplicity notion} (membership in~$\np$).

Thus we redefine resistance to mean $\np$-hardness.  This redefinition
was already praised, though not adopted, 
in~\cite[Footnote~9]{hem-hem-rot:j:destructive-control}.  
Note that all existing resistance results (under the old definition)
imply resistance under the new definition---which should lessen the
potential for confusion.

As a final comment on this redefinition, we above casually mentioned
that control problems might in difficulty exceed $\np$-completeness.
But can that really occur in truly natural systems?
The answer is yes.  In particular, note that the unique winner problem
for any election system (with preference orders) reduces to its
constructive control by adding candidates problem (by setting the
spoiler candidate set $D$ to~$\emptyset$), to its
constructive control by deleting candidates problem (by setting the
deletion limit $k$ to~$0$), to its
constructive control by adding voters problem (by setting the
addition bound $k$ to~$0$), and to its
constructive control by deleting voters problem (by setting the
deletion limit $k$ to~$0$).
Thus \emph{every} system with a hard winner problem has four control
problems that are at least as hard.
Note that Dodgson~\cite{dod:unpub:dodgson-voting-system}, 
Kemeny~\cite{kem-sne:b:polsci:mathematical-models}, and
Young~\cite{you:j:extending-condorcet,lev-you:j:condorcet}
elections are much-studied natural systems whose winner problem
(but defined in terms of ``winner,'' not ``unique winner'')
is known---respectively due
to~\cite{hem-hem-rot:j:dodgson},
\cite{hem-spa-vog:j:kemeny}, and
\cite{rot-spa-vog:j:young}---to
be complete for the $\Theta_{2}^{p}$
level~\cite{pap-zac:c:two-remarks,wag:j:bounded} of the polynomial
hierarchy (a level that contains~$\np$, $\conp$, and the boolean
hierarchy~\cite{cai-gun-har-hem-sew-wag-wec:j:bh1}, so if even one
$\Theta_{2}^{p}$-hard set is $\np$-complete then the polynomial
hierarchy collapses to~$\np$).
From this and the above comments, it follows that the four mentioned
control problems are $\Theta_{2}^{p}$-hard
for each of Dodgson, Kemeny, and Young
elections when the control problems are altered to speak of ``winner''
rather than ``unique winner'' (this alteration is \emph{not} the model
of this paper or any previous paper on control).
However, 
we observe by close inspection that in all of
\cite{hem-hem-rot:j:dodgson,hem-spa-vog:j:kemeny,rot-spa-vog:j:young}, the
proofs that establish $\Theta_{2}^{p}$-hardness for the Dodgson, Kemeny,
and Young election winner problems can be modified to show that their
\emph{unique} winner problems are also 
$\Theta_{2}^{p}$-hard.\footnote{As mentioned above, the winner problems 
for Dodgson~\cite{dod:unpub:dodgson-voting-system}, 
Kemeny~\cite{kem-sne:b:polsci:mathematical-models}, and
Young~\cite{you:j:extending-condorcet,lev-you:j:condorcet}
elections are known---respectively due
to~\cite{hem-hem-rot:j:dodgson}, \cite{hem-spa-vog:j:kemeny},
and \cite{rot-spa-vog:j:young}---to
be complete for~$\Theta_{2}^{p}$,
the complexity class capturing ``parallel access to~$\np$.''
In this footnote, we show the analogous results for the unique
winner problems. (See the above papers and~\cite{bar-tov-tri:j:who-won}
for the definitions
of these three election systems and of Dodgson score.)

\begin{theorem}
\label{t:dodgson-and-others}
The unique winner problem for Dodgson elections is $\thetatwo$-complete.
The unique winner problem for Kemeny elections is $\thetatwo$-complete.
The unique winner problem for Young elections is $\thetatwo$-complete.
\end{theorem}

\begin{proofs}
$\thetatwo$-completeness of the winner problem for Dodgson elections
was shown in~\cite{hem-hem-rot:j:dodgson}.  The $\thetatwo$ upper bound
is easy to show, and the same argument can be used to show a 
$\thetatwo$ upper bound for the unique winner problem.  It remains
to show that the unique winner problem for Dodgson elections is
$\thetatwo$-hard. $\thetatwo$-hardness of the winner problem
for Dodgson elections was shown 
by a reduction from the $\thetatwo$-hard problem 
Two-Election Ranking (2ER, for short)~\cite{hem-hem-rot:j:dodgson}:
Given two Dodgson triples $\parpair{C,c,V}$ and $\parpair{D,d,W}$, where
both $\|V\|$ and $\|W\|$ are odd and $c \neq d$,
is the Dodgson score of $c$ in $\parpair{C,V}$ less than or equal to 
the Dodgson score of $d$ in $\parpair{D,W}$?
A Dodgson triple $\parpair{C,c,V}$ is an election $\parpair{C,V}$
with a designated candidate $c \in C$.

The reduction from 2ER to the winner problem for
Dodgson elections~\cite{hem-hem-rot:j:dodgson} has the property
that if an instance $I = \parpair{\parpair{C,c,V},\parpair{D,d,W}}$
of 2ER is mapped to the Dodgson triple $(\widehat{C},c,\widehat{V})$, then 
$c$ and $d$ are winners of $(\widehat{C},\widehat{V})$ if
$I$ is in 2ER, and $d$ is the unique winner of $(\widehat{C},\widehat{V})$
if $I$ is not in 2ER.  It follows that $I$ is in 2ER if and only
if $d$ is not the unique Dodgson winner of $(\widehat{C},\widehat{V})$.
Thus the complement of the unique winner problem
for Dodgson elections is $\thetatwo$-hard.  Since $\thetatwo$ is closed
under complement, the unique winner problem
for Dodgson elections is also 
$\thetatwo$-hard.
$\thetatwo$-completeness of the winner problem for Kemeny elections
was shown in~\cite{hem-spa-vog:j:kemeny}.  The $\thetatwo$ upper bound
is easy to show, and the same argument can be used to show a 
$\thetatwo$ upper bound for the unique winner problem.  It remains
to show that the unique winner problem for Kemeny elections is
$\thetatwo$-hard. $\thetatwo$-hardness of the winner problem
for Kemeny elections was shown 
by a reduction from the $\thetatwo$-hard problem Feedback Arc Set
Member~\cite{hem-spa-vog:j:kemeny}:
Given an irreflexive 
and antisymmetric 
digraph $G$ and a vertex $c$ of $G$,
does there exist a minimum-size feedback arc set of $G$ that contains
all arcs entering $c$?  A feedback arc set of $G$ is a set of arcs
that contains at least one arc from every cycle in $G$.

The reduction from Feedback Arc Set Member to 
the winner problem for Kemeny elections~\cite[Lemma 4.2]{hem-spa-vog:j:kemeny}
works as follows.  There is a polynomial-time computable function $g$
such that for all irreflexive and antisymmetric digraphs $G$,
$g(G)$ is an election whose candidates are
the vertices of $G$ and such that for all vertices $c$ of $G$,
$\pair{G,c}$ is in Feedback Arc Set Member if and only if $c$ is a
Kemeny winner of election $g(G)$.

We turn this into a reduction from the complement of
Feedback Arc Set Member to the unique winner problem for Kemeny elections
as follows:  Given an input instance $\pair{G,c}$ of 
Feedback Arc Set Member, add a new vertex $d$ to $G$ and
add arcs from $d$ to every vertex in $V(G) - \{c\}$.
Note that no minimum-size feedback arc set of $\widehat{G}$ contains
an arc starting from $d$ (since these arcs are not part of a cycle), and so 
for all $c' \in V(G) - \{c\}$, $\pair{\widehat{G},c'}$ is
not in Feedback Arc Set Member.  Also note that $\pair{\widehat{G},d}$ is
trivially in Feedback Arc Set Member since there are no arcs entering $d$.  

Now consider election $g(\widehat{G})$. Recall that the candidates of 
$g(\widehat{G})$ are the vertices of $\widehat{G}$ and that for all
vertices $c'$ in $\widehat{G}$, $\pair{\widehat{G},c'}$ is in
Feedback Arc Set Member if and only if $c'$ is a Kemeny winner of election
$g(\widehat{G})$.  From the observations above, it follows immediately
that $\pair{\widehat{G},c}$ is in Feedback Arc Set Member if and only
if $d$ is not the unique Kemeny winner of election $g(\widehat{G})$. Since
$\pair{G,c}$ is in Feedback Arc Set Member if and only if
$\pair{\widehat{G},c}$ is in Feedback Arc Set Member,
we have obtained a polynomial-time reduction from 
Feedback Arc Set Member to the complement of the
unique winner problem for Kemeny elections. 
Since $\thetatwo$ is closed under complement, this proves that
the unique winner problem for Kemeny elections is 
$\thetatwo$-hard.
$\thetatwo$-completeness of the winner problem for Young elections
was shown in~\cite{rot-spa-vog:j:young}.
The $\thetatwo$ upper bound
is easy to show, and the same argument can be used to show a 
$\thetatwo$ upper bound for the unique winner problem.
$\thetatwo$-hardness of the winner problem
for Young elections was shown as Theorem~2.5 
in~\cite{rot-spa-vog:j:young} via a reduction from the
$\thetatwo$-complete problem Maximum Set Packing Compare (MSPC,
for short)~\cite{rot-spa-vog:j:young}: Given two sets, $B_1$ 
and~$B_2$, and two collections of sets, $\mathcal{S}_1$ 
and~$\mathcal{S}_2$, where each $S \in \mathcal{S}_i$ is a
nonempty, finite subset of~$B_i$, is it the case that
$\kappa(\mathcal{S}_1) \geq \kappa(\mathcal{S}_2)$?
Here, $\kappa(\mathcal{S}_i)$ denotes the maximum number of pairwise
disjoint sets in~$\mathcal{S}_i$.

The proofs of Theorems~2.3 and~2.5 in~\cite{rot-spa-vog:j:young}
show how to construct from a
given MSPC instance
$I = \parpair{B_1, B_2, \mathcal{S}_1, \mathcal{S}_2}$ an election 
$\parpair{C,V}$ with two designated candidates, $c$ and~$d$, such that
(a) if $\kappa(\mathcal{S}_1) \geq \kappa(\mathcal{S}_2)$ then 
$c$ and $d$ are Young winners of $({C},{V})$, and (b) if
$\kappa(\mathcal{S}_2) > \kappa(\mathcal{S}_1)$ then 
$d$ is the unique Young winner of $({C},{V})$.
It follows that $I$ is in MSPC if and only
if $d$ is not the unique Young winner of $({C},{V})$.
Thus the complement of the unique winner problem
for Young elections is $\thetatwo$-hard. 
Since $\thetatwo$ is closed under complement, this proves that
the unique winner problem for Young elections is 
$\thetatwo$-hard.~\end{proofs}
}
Thus, we may state the following.

\begin{theorem}
\label{t:unique-winner-dodgson-kemeny-young}
For each of Dodgson elections, Kemeny elections, and Young elections,
the following control problems are $\Theta_{2}^{p}$-hard:
constructive control by adding candidates,
constructive control by deleting candidates,
constructive control by adding voters, and
constructive control by deleting voters.
\end{theorem}

An anonymous 
referee commented that
even polynomial-time algorithms can be expensive to run on
sufficiently large inputs.
We mention that though the comment is correct, almost any
would-be controller would probably much prefer that challenge,
solving a P problem on large inputs, to the challenge our
results give him/her, namely, solving an NP-complete problem on large
inputs.  We also mention that since
the hybrid scheme is designed so as to inherit resistances from
the underlying schemes, if a hybrid requires extreme ratios
between the number of candidates and the number of voters to display
asymptotic hardness, that is purely due to inheriting that from
the underlying systems.  Indeed, if anything the hybrid is less
likely to show that behavior since, informally put, if even one
of the underlying systems achieves asymptotic hardness even
away from extreme ratios between the number of candidates and the number
of voters, then their hybrid will also.

\subsection{Inheritance}

We will be centrally concerned with the extent to which
$\hybrid(\election_0, \election_1, \ldots , \election_{k-1})$
inherits the properties of its constituents.  To do so, we formally
define our notions of inheritance (if all the constituents have a
property then so does their hybrid) and of strong inheritance (if even
one of the constituents has a property then so does the hybrid).

\begin{definition}
\label{d:inherit}
We say that a property $\Gamma$
is \emph{strongly inherited} (respectively,
\emph{inherited}) by $\hybrid$ if
 the following holds:
Let $k \in \nats^+$.  Let $\election_0, \election_1, \ldots ,
\election_{k-1}$ be candidate-anonymous election systems (each taking
as input $\parpair{C,V}$, with $V$ a list of preference orders).  It holds that
$\hybrid(\election_0, \election_1, \ldots , \election_{k-1})$ has property
$\Gamma$ if at least one of its constituents has (respectively, all its
constituents have) property~$\Gamma$.
\end{definition}

Definition~\ref{d:inherit} builds in the assumption that all constituents
are candidate-anonymous.  This assumption isn't overly onerous since as
mentioned earlier candidate-anonymity is very common---but will be used
in many of our proofs.  In effect, it is the price we pay for our choice
of ``switching between systems'' rule within $\hybrid$.  One might
naturally worry that as $\hybrid$ is not itself candidate-anonymous,
inheritance results might be difficult to prove for a hybrid of hybrids.
That is true, but before reaching such an impasse one would typically avoid it,
namely, instead of trying to study
\[
\hybrid(\hybrid(\election_0, \election_1, \ldots ,
\election_{k-1}),\hybrid(\election_0', \election_1', \ldots ,
\election_{k'-1}'))
\]
rather studying $\hybrid(\election_0, \election_1, \ldots ,
\election_{k-1},\election_0', \election_1', \ldots , \election_{k'-1}')$.
Though there of course are issues regarding default handling, in most settings
this will not have problems, and in fact we will do precisely this later
(in the proof of Theorem~\ref{t:all-twenty}).  

Though we will build candidate-anonymity into the assumptions underlying
inheritance, we will often try to let interested readers know when that
assumption is not needed.  In particular, when we say ``inherited (and
flexibly so)'' or ``strongly inherited (and flexibly so),'' the ``(and
flexibly so)'' indicates that the claim holds even if in
Definition~\ref{d:inherit} the words ``candidate-anonymous'' are deleted.  For
example, the following easy but quite important claim follows easily from the
definition of $\hybrid$.
\begin{proposition}
\label{t:winner-p}
``Winner problem membership in $\p$''  and
``unique winner problem membership in $\p$'' are
 inherited (and flexibly so) by
$\hybrid$.
\end{proposition}
That is, if $\election_0, \election_1, \ldots , \election_{k-1}$ are election
rules (all, as we assume globally, based on input preference orders) for which
testing whether a given candidate is the winner is in $\p$ (i.e., 
$\{\parpair{C,V,c}
\condition \mbox{$c$ is a winner in the election 
specified by $\parpair{C,V}$} \} \in \p$), 
then the winner problem for $\hybrid(\election_0, \election_1,
\ldots , \election_{k-1})$ is also in~$\p$.  We similarly have that the
following claim follows easily from the definition of $\hybrid$.
\begin{proposition}
\label{t:winner-np}
``Winner problem membership in $\np$'' and
``unique winner problem membership in $\np$'' are
inherited (and flexibly so) by
$\hybrid$.
\end{proposition}
(In fact, it is easy to see that when the constituent systems are
candidate-anonymous then, for each $k \geq 0$, it holds that 
the winner problem (respectively, unique winner problem)
of $\hybrid$ is in $\Sigma_k^p$ if and only if
all the constituent winner problems (respectively, unique winner problems)
are in $\Sigma_k^p$, where $\Sigma_k^p$ as is
standard denotes the $k$th level of the polynomial hierarchy.)

On the other hand, sometimes we will want to focus just on elections 
having a certain property $\Phi$, and for those we use the following
definition.

\begin{definition}
\label{d:inherit-on}
We say that a property $\Gamma$
is \emph{strongly inherited} (respectively,
\emph{inherited}) \emph{by $\hybrid$ on election systems having property $\Phi$}
 if the following holds:
Let $k \in \nats^+$.  Let $\election_0, \election_1, \ldots ,
\election_{k-1}$ be candidate-anonymous election systems (each taking
as input $\parpair{C,V}$, with $V$ a list of preference orders)
each having property $\Phi$.  It holds that
$\hybrid(\election_0, \election_1, \ldots , \election_{k-1})$ has property
$\Gamma$ if at least one of its constituents has (respectively, all its
constituents have) property~$\Gamma$.
\end{definition}

\subsection{Discussion of Some Concrete Objections}
It is important when presenting an election system to make 
clear what the purpose of defining it is and how it 
is intended to be used.   This was already discussed 
in part in 
Sections~\ref{s:intro} 
and~\ref{ss:scheme}.
However, we now have defined for the reader 
our hybridization scheme and key notions 
such as control and inheritance, and so we are in a better 
position to 
be able to 
address potential worries.  
So, as a way both of recapping some of 
the points of 
Sections~\ref{s:intro} 
and~\ref{ss:scheme}
and of 
making some additional important points about 
the nature of our hybridization system, we will now 
state and discuss some interesting
issues that one might wonder
about.

One natural issue is that our model is about worst-case hardness
and so may not speak to typical (frequent) hardness.  One 
anonymous referee
observed that since the vast majority of the inputs will have mismatches in
the moduli 
of the names 
(i.e., will have 
mismatches in what the input names are congruent to modulo $k$) 
and so will map to the default system, 
the election systems other than the default one ``do not really matter.''

This actually raises a number of issues.  We will try to 
now address them briefly but clearly.

First, and most 
importantly, the core purpose of the hybrid scheme is not to be a system
that is used directly as a practical, real-world 
election system, but rather is, as discussed in 
the introduction, to show that the impossibility result:
\begin{quote}
``For no election system whose winner complexity is in P             
  are all twenty types of control $\np$-hard.''
\end{quote}
outright does not hold.  And by showing that we can combine control
resistances while not boosting the complexity of the winner problem beyond
P, our hybrid scheme in fact breaks all hope of such an impossibility
result, via showing that there is an election system whose winner
complexity is in P yet for which all twenty types 
of control
are $\np$-hard (Theorem~\ref{t:all-twenty}).  

Second, the hybrid scheme's point is to in effect ``join'' its
underlying systems.  It provides a one-stop system that allows one (by
appropriately renaming the candidates) to get the effect of any one of
the underlying systems (assuming candidate-anonymity holds for 
those).  As to the referee's comment that the vast majority of the
inputs will have mismatches in the moduli of the names and so will
map to the default system, and so the election 
systems other than the default one ``do not really
matter,'' the comment works best if 
one views the entire hybrid as a
stand-alone system intended for direct, natural use, and if one further 
feels that such
use should be with respect to a uniform distribution of inputs directly
to $\hybrid$.
But as argued in the previous paragraph, the hybrid scheme is not intended as
a system for direct, natural, real-world use; it is built
to give insight into inheritance of control resistance and 
most critically into
(non)existence of impossibility results.  Further, even if one did
want to look at the system relative to a uniform notion of inputs,
given that the system (as mentioned at
the start of the present paragraph) is
intended to combine $k$ systems, the natural notion of a uniform input
draw would be to randomly choose one of the $k$ systems, randomly
generate an instance there, and then mod-encode the names to place
that instance into the hybrid scheme.  Note that in this
approach/distribution, the frequency of having inputs route to the
default system due to name moduli mismatches would not be very
frequent but would be ``absolutely never.''  (As a very loose 
and admittedly imperfect analogue,
if one types random keystrokes, one is very unlikely to get 
a Boolean formula that is syntactically well-formed.  And so 
relative to an utterly uniform distribution of that sort, 
SAT becomes trivial in the vast majority of cases.  For us, 
agreement of name moduli is a form of syntactic filter used to 
discipline the input, and 
inputs lacking agreement are swept to the default system.)

\clearpage
Third, let us make clear that there is a very real, important issue in
the concrete 
objection being 
discussed.  Namely, what about frequency of hardness, or
even about average-case hardness?  Clearly, the 
hypothetical impossibility theorem above 
itself includes the notion of NP-completeness, which is a worst-case
theory, rather than one focused on either average-case NP-hardness or
frequency of hardness.  It is indeed 
very natural to hope to achieve
average-case NP-hardness or frequency of hardness results.  We fully
agree that 
this is a very interesting, important 
long-term issue.  But we briefly mention
some cautions and limitations.  Regarding average-case NP-hardness,
a well-known theory developed by Levin~\cite{lev:j:average-case},
the theory is lovely, but only a handful of problems have been shown
to be average-case NP-hard, and not a single one of those in any way
involves elections.  Regarding frequency of hardness, there have been
a number of recent results showing that certain 
NP-hard election issues can be
frequently simple, but to the best 
of our knowledge, none of those papers obtains such results for 
any control problem.\footnote{All such work 
that we know of regards winner and 
manipulation problems, not control problems, and even 
for the winner and control cases such results typically require one to 
choose the distribution very carefully.
For example, the work of
\cite{hem-hom:cOutByJToAppear:dodgson-greedy,hem-hom:jtoappear:dodgson-greedy}
and
\cite{mcc-pri-sli:j:approximability-of-dodgson}
succeeds in the uniform distribution, 
but is not known to hold under other
distributions, and anyway
is about winner problems and 
not about control or manipulation.
And in the interesting 
work of \cite{con-san:c:nonexistence},
which is about manipulation (not about control), 
the frequency of being able to manipulate is conditioned inside the
set of cases that are manipulable at all.  For the one
concrete distribution family on which that paper does experiments, it
is easy to see that except for the special case of the uniform
distribution, ``$p=0.5$'' in their notation, asymptotically the
portion of cases that are manipulable is extraordinarily small
(because whomever is at the top of what they call the ``correct''
ranking will as the number of voters grows, win by a large amount when
$p\neq 0.5$).  So what one should take away is 
not that 
the election systems are almost always easily manipulable, but rather
that the election system
they studied are, for that distribution, almost never manipulable (but
nonetheless that, and this is that paper's remarkable insight, 
conditioning within 
the extraordinarily low probability-weight of instances in which they are
manipulable, one can with high probability easily find a manipulation,
again with this all framed within that distribution).  Without
trying to here list all the other interesting papers addressing 
issues of frequency of correctness, as there are many and all
of them are not about the present paper's focus (namely, control), 
we mention as particularly interesting 
(and provocative) the work of 
Procaccia and Rosenschein on juntas
and manipulation~\cite{pro-ros:j:juntas},
the 
Procaccia and Rosenschein fraction-of-manipulators 
work~\cite{pro-ros:c:average-case-fraction-of-manipulators},
the work of Friedgut, Kalai, and Nisan
on random manipulations~\cite{fri-kal-nis:ctoappear:manipulated-often},
and the Zuckerman, Procaccia, and Rosenschein work on
coalition manipulation~\cite{pro-ros-zuc:c:coalition-manipulation}.
}
Briefly summarized, issues of average-case
NP-hardness and of frequency of correctness are clearly very
interesting, are 
rightly gaining attention, 
and we highly commend them to the reader while also mentioning that 
the field's
understanding of such results and ability to obtain them is still
very limited
and currently doesn't exist at all for the case of control problems
(except see the comment of the next paragraph, which in a sense 
says that our work at least preserves good behavior in this regard).

Still continuing on this third point, however, namely of frequency of
hardness, in the particular case of the present paper's work we
\emph{do} have a case where---in a certain rigorous sense that is theoretically
crisp although practically quite unsatisfying---inheritance of
hardness does
hold.  In
particularly, it is clear, directly from our definitions, that
hardness inherits under our scheme---indeed, that hardness strongly
inherits under our scheme.  In particular, if even just one of the
constituent systems (all required to be candidate anonymous) is
exponentially often exponentially hard, then so is the hybridization
of those systems.  The catch here is that it is a ``small''
exponential regarding the frequency of hardness, namely
$2^{\Omega(n)}$, and so it ensures an exponentially rich core of hard
instances, but it still could be the case that the overwhelming
majority of instances are easy.\footnote{For completeness, let us be
  specific as to what is being claimed.  Let us say a problem $A$ is
  exponentially often exponentially hard if there is an $\epsilon > 0$
  such that, for each Turing machine $M$ accepting $A$, there are an
  infinite number of lengths $n$ at which $M$ runs for at least
  $2^{\epsilon n}$ steps on at least $2^{\epsilon n}$ of the strings
  of that length.  Our theorem is this: Being exponentially often
  exponentially hard is strongly inherited (in the sense of
  Definition~\ref{d:inherit}) by $\hybrid$.  This is true simply
  because the process of taking an instance from one of the
  constituent elections and recoding the names in the natural 
  way to recast the instance into our
  hybrid scheme increases the instance's length at most linearly.
  
  Our framing of this claim just said ``there are an infinite number 
  of lengths'' at which the hardness holds, but note that, again 
  due to the fact that we can achieve
  at-most-linear instance-length-stretching, whatever the original 
  density of hard lengths is of the hard constituent system, 
  the density of hard lengths in $\hybrid$ will be no more than 
  linearly less.  Informally put,
  we don't lose too much density of hard lengths.

  We mention in passing that the above claim remains true if each
  $2^{\Omega(n)}$ above is changed to $2^{n^{\Omega(1)}}$, a different,
  even smaller notion of having exponentially many elements.  Although
  $2^{\Omega(n)}$ and $2^{n^{\Omega(1)}}$ are not practically very
  satisfyingly large when compared (regarding frequency) with $2^n$, we
  mention that in theoretical computer science 
  this type of density is very natural
  and typical in such settings.
  This is due to the fact that for each $\epsilon > 0$, every NP-hard
  set $A$ is, by standard padding arguments, known to be
  polynomial-time many-one equivalent to sets (e.g., $\{0^{|x|^k}\#x
  \mid x \in A\}$, where $k$ is chosen to satisfy $k > 1/\epsilon$)
  that for each $n$ are trivial on all but $2^{n^{\epsilon}}$ of their
  length $n$ strings.  Indeed, for almost precisely this reason,
  Levin's theory of average-case NP-hardness~\cite{lev:j:average-case} 
  allows some sets that are
  trivial on all but at most $2^{n^{\epsilon}}$ of their length $n$ strings to
  be average-case NP-hard.
}

However, we should not allow ourselves to be overly sidetracked.  The
main points here are simply that our hybrid system is primarily designed
to show that control can be strongly inherited, and it does so and
thus completely rules out the natural, potential impossibility result.
And that issues of frequency of hardness are indeed interesting,
although difficult and typically very distribution-based, and deserve
(and are getting) further investigation.  Regarding the first of these
points, one might ask whether one really needs an artificial system to
preclude the impossibility result.  The authors and P.~Faliszewski
have, motivated by the present paper, 
looked hard at this issue.  As mentioned earlier, 
they have very recently shown
(in \cite{fal-hem-hem-rot:c:llull} give or take one's view on a
technical fine point, and in \cite{fal-hem-hem-rot:c:copeland-alpha}
regardless of one's view on the technical fine point) that there is a
natural system---one of the versions of Copeland voting---that has a P
winner problem yet is immune to all ten types of constructive control.
However, that system is vulnerable to most types of destructive
control.  And so their work gives an alternate path to 
precluding
a ten-constructive-case
impossibility result, but unlike the present paper does not preclude
the main goal---a twenty-case overall impossibility result---or
even preclude a
ten-destructive-case impossibility result.

This completes our discussion regarding whether systems other than the
default really matter in $\hybrid$, and of issues of average-case
NP-hardness and frequency of hardness.  We now turn to a second
concrete worry one might have.  

One anonymous referee 
proposed a different hybrid scheme than ours based on, on
each input, randomly choosing a system to route the question to.
Unfortunately, that approach to a hybrid scheme is not a valid
election rule, as it uses probability not just to compute something,
but also its output itself is probabilistic.  That is, if one were to
repeatedly run the exact same set of completed ballots through that
``rule,'' one would get different winners on different such runs.
That isn't an election rule in the standard sense universally used to
study control, and the reduction framework (that we ourselves exploit
to prove hardness)
does not apply to this.  In addition, the present paper \emph{proves}
results for its system.

We now turn to a third concrete worry.  Our paper's key tool is a
hybrid scheme.  We then apply it using certain building blocks
(constituent election systems to be hybridized together).  Let us say
a voting system is \emph{voiced}
(see~\cite{hem-hem-rot:j:destructive-control}) if (whenever there is
at least one candidate) there will be at least one winner.  Although
nonvoiced systems have been allowed in the study of control in every
paper we know of on control, one might dislike the fact that in
some places in this paper we use building blocks that are not voiced
(may sometimes have no winner).  The two building blocks we use that
are not voiced are Condorcet elections and an artificial election
system we call $\electionnotallone$.  However, we point out that one
can just as easily toss out those two building blocks and use instead
any other existing system(s) that have the resistance properties that
the discarded blocks were being used for and that in addition are
voiced.  Such systems have been proven 
(although just
recently)
to exist.  Combined with the easy-to-see claim that
voiced-ness inherits under our hybrid scheme, one can, if one dislikes
nonvoiced systems, establish the theorems in question via a voiced
hybrid of voiced systems.\footnote{To be utterly concrete about this,
Condorcet voting is used in this paper for its resistance to constructive
control by adding, deleting and partitioning of voters, and
$\electionnotallone$ is used in this paper for its resistance to
destructive control by adding, deleting and partitioning of voters (in
the TE model).  However, very recent
work~\cite{fal-hem-hem-rot:c:llull} has proven that the natural, voiced
election system of Llull (and also the natural, voiced 
election system of
Copeland and many
variants~\cite{fal-hem-hem-rot:c:llull,fal-hem-hem-rot:c:copeland-alpha})
itself has all of these resistance properties, and so one can, 
if one requires voiced systems, employ Llull elections here in place 
of both 
Condorcet
and $\electionnotallone$.}

\section{Inheritance and Hybrid Elections: 
Susceptibility, Resistance, Immunity, and Vulnerability}
\label{s:sriv}

In this section we will discuss the inheritance properties of $\hybrid$ with
respect to susceptibility, resistance, immunity, and vulnerability.
Tables~\ref{tab:constructive}
and~\ref{tab:destructive} summarize our results
for the cases of, respectively, constructive control and destructive
control.
(These tables do not discuss/include the issue of when ``(and flexibly so)''
holds, i.e., when the candidate-anonymity assumption is not needed, but rather
focus on 
our basic inheritance definition.)

\begin{table*}[tp]
\small
\centering
\begin{tabular}{||l||c|c|c|c||}
\hline\hline
Control by & Susceptibility & Resistance & Immunity & Vulnerability \\ 
\hline\hline
Adding Candidates   & SI & SI & Not I & I \\ 
 & \ref{c:ABK} &
   \ref{c:r-gate}  & \ref{t:immunity-not-inherited-results} &
   \ref{t:vulnerability-inherited-for-adding-candidates} \\ \hline
Deleting Candidates & SI & SI & I     & I if and only if $\p = \np$ \\ 
 & \ref{c:ABK} &
   \ref{c:r-gate}  & 
   \ref{t:hybrid-inherits-immunity-to-constr-control-by-deleting-candidates} &
   \ref{c:characterization-constr-control-deleting-cand} \\ \hline
Partition           & SI & SI & Not I & On election systems having\\
of Candidates (TE)  &    &    &       & unique winner problems in PH:\\
& & & & I iff SI iff $\p = \np$\\
 & \ref{c:ABK} &
   \ref{c:r-gate}  & \ref{t:immunity-not-inherited-results} &
\ref{c:characterization-constr-control-partition-cand}
   \\ \hline
Partition           & SI & SI & Not I & On election systems having\\
of Candidates (TP)  &    &    &       & winner problems in PH:\\
& & & & I iff SI iff $\p = \np$\\
 & \ref{c:ABK} &
   \ref{c:r-gate}  & \ref{t:immunity-not-inherited-results} &
\ref{c:characterization-constr-control-partition-cand}
   \\ \hline
Run-off Partition           & SI & SI & Not I & On election systems having\\
of Candidates (TE)  &    &    &       & unique winner problems in PH:\\
& & & & I iff SI iff $\p = \np$\\
 & \ref{c:ABK} &
   \ref{c:r-gate}  & \ref{t:immunity-not-inherited-results} &
\ref{c:characterization-constr-control-run-off-part-cand}
   \\ \hline
Run-off Partition           & SI & SI & Not I & On election systems having\\
of Candidates (TP)  &    &    &       & winner problems in PH:\\
& & & & I iff SI iff $\p = \np$\\
 & \ref{c:ABK} &
   \ref{c:r-gate}  & \ref{t:immunity-not-inherited-results} &
\ref{c:characterization-constr-control-run-off-part-cand}
   \\ \hline
Adding Voters       & SI & SI & I     & I \\ 
 &  \ref{c:ABK} &
    \ref{c:r-gate}  &
   \ref{t:eight-hybrid-inherits-immunity-results} &
   \ref{t:vulnerability-inherited-for-voter-control} \\ \hline
Deleting Voters     & SI & SI & I     & I \\ 
 &  \ref{c:ABK} &
   \ref{c:r-gate}  &
   \ref{t:eight-hybrid-inherits-immunity-results} &
   \ref{t:vulnerability-inherited-for-voter-control} \\ \hline
Partition           & SI & SI & I     & I \\
of Voters (TE)      & \ref{c:ABK} &
  \ref{c:r-gate}     &
   \ref{t:eight-hybrid-inherits-immunity-results} &
   \ref{t:vulnerability-inherited-for-voter-control} \\ \hline
Partition           & SI & SI & I     & I \\ 
of Voters (TP)      &  \ref{c:ABK} &
  \ref{c:r-gate}    &
   \ref{t:eight-hybrid-inherits-immunity-results} &
   \ref{t:vulnerability-inherited-for-voter-control} \\ \hline\hline
\end{tabular}
\caption{Inheritance results that hold or provably fail for $\hybrid$ with
  respect to constructive control.  Key: I = Inherits. SI = Strongly Inherits.
  The numbers in each box give results establishing or implying that box's
  claim.
   Those table entries in the vulnerability column containing
  ``on election systems'' clauses are somewhat informally stated above; see the
  cited results for precise statements.
\label{tab:constructive}
}
\end{table*}

\begin{table*}[tp]
\small
\centering
\begin{tabular}{||l||c|c|c|c||}
\hline\hline
Control by & Susceptibility & Resistance & Immunity & Vulnerability \\
\hline\hline
Adding Candidates   & SI & SI & I     & I \\ 
 & \ref{c:ABK} &
   \ref{c:r-gate}  &
   \ref{c:hybrid-inherits-immunity-to-destr-control-by-adding-candidates} &
   \ref{t:vulnerability-inherited-for-adding-candidates} \\ \hline
Deleting Candidates & SI & SI & Not I & I if and only if $\p = \np$ \\
 & \ref{c:ABK} &
   \ref{c:r-gate}  & \ref{t:immunity-not-inherited-results} &
   \ref{c:characterization-destr-control-deleting-cand} \\ \hline
Partition           & SI & SI & Not I & On election systems having\\
of Candidates (TE)  &    &    &       & unique winner problems in PH:\\
& & & & I iff SI iff $\p = \np$\\
 & \ref{c:ABK} &
   \ref{c:r-gate}  & \ref{t:immunity-not-inherited-results} &
\ref{c:characterization-destr-control-partition-cand}
   \\ \hline
Partition           & SI & SI & Not I & On election systems having\\
of Candidates (TP)  &    &    &       & winner problems in PH:\\
& & & & I iff SI iff $\p = \np$\\
 & \ref{c:ABK} &
   \ref{c:r-gate}  & \ref{t:immunity-not-inherited-results} &
\ref{c:characterization-destr-control-partition-cand}
   \\ \hline
Run-off Partition           & SI & SI & Not I & On election systems having\\
of Candidates (TE)  &    &    &       & unique winner problems in PH:\\
& & & & I iff SI iff $\p = \np$\\
 & \ref{c:ABK} &
   \ref{c:r-gate}  & \ref{t:immunity-not-inherited-results} &
\ref{c:characterization-destr-control-partition-cand}
   \\ \hline
Run-off Partition           & SI & SI & Not I & On election systems having\\
of Candidates (TP)  &    &    &       & winner problems in PH:\\
& & & & I iff SI iff $\p = \np$\\
 & \ref{c:ABK} &
   \ref{c:r-gate}  & \ref{t:immunity-not-inherited-results} &
\ref{c:characterization-destr-control-partition-cand}
   \\ \hline
Adding Voters       & SI & SI & I     & I \\ 
 & \ref{c:ABK} &
   \ref{c:r-gate}  &
   \ref{t:eight-hybrid-inherits-immunity-results} &
   \ref{t:vulnerability-inherited-for-voter-control} \\ \hline
Deleting Voters     & SI & SI & I     & I \\ 
 & \ref{c:ABK} &
   \ref{c:r-gate}  &
   \ref{t:eight-hybrid-inherits-immunity-results} &
   \ref{t:vulnerability-inherited-for-voter-control} \\ \hline
Partition           & SI & SI & I     & I \\
of Voters (TE)      & \ref{c:ABK} &
  \ref{c:r-gate}   &
   \ref{t:eight-hybrid-inherits-immunity-results} &
   \ref{t:vulnerability-inherited-for-voter-control} \\ \hline
Partition           & SI & SI & I     & I \\ 
of Voters (TP)      & \ref{c:ABK} &
  \ref{c:r-gate}   & 
   \ref{t:eight-hybrid-inherits-immunity-results} &
   \ref{t:vulnerability-inherited-for-voter-control} \\ \hline\hline
\end{tabular}
\caption{Inheritance results that hold or provably fail for $\hybrid$ with
  respect to destructive control.  Key: I = Inherits. SI = Strongly Inherits.
  The numbers in each box give results establishing or implying that box's
  claim.
   Those table entries in the vulnerability column containing
  ``on election systems'' clauses are somewhat informally stated above; see the
  cited results for precise statements.
\label{tab:destructive}
}
\end{table*}

\subsection{Susceptibility}

We first note that susceptibility strongly inherits.  In fact, we prove
something slightly stronger, as for only the susceptible constituent do we
need to assume candidate-anonymity (also, as noted in the introduction, 
candidate-affine-invariance would actually suffice).
We remind the reader that throughout this paper, when we speak of an election
system, we always implicitly mean an election system that takes as input
$\pair{C,V}$ (see Footnote~\ref{f:88-candidate-argument}) with $V$ a list of
preference orders over~$C$.

\begin{theorem}
\label{t:ABK}
Let $k \in \nats^+$ and let $\election_0, \election_1, \ldots ,
\election_{k-1}$ be election systems.  Let $\Phi$ be one of the standard
twenty types of (constructive and destructive) control.  If for at least
one~$i$, $0 \leq i \leq k-1$, $\election_i$ is candidate-anonymous and
susceptible to~$\Phi$, then $\hybrid(\election_0, \election_1, \ldots ,
\election_{k-1})$ is susceptible to~$\Phi$.
\end{theorem}

\begin{proofs}
  Take an
 example showing that $\election_i$ is susceptible to $\Phi$ and
  rename each candidate $c$ in it to now be named $kc+i$.  This transformed
  example will, due to $\election_i$'s candidate-anonymity, be an example
  showing $\hybrid(\election_0, \election_1, \ldots , \election_{k-1})$'s
  susceptibility to~$\Phi$.~\end{proofs}

We note in passing that there is nothing particularly limiting the above
theorem to the standard twenty types of control.  The embedding would work
equally well on almost any natural control type one might imagine.

Theorem~\ref{t:ABK} immediately yields the following corollary.

\begin{corollary}
\label{c:ABK}
$\hybrid$ strongly inherits susceptibility to each of the standard twenty
types of
control.
\end{corollary}

\subsection{Resistance}
\label{sec:resistance}

We now come to the most important inheritance case, namely, that of
resistance.  Since our hope is that hybrid elections will broaden resistance,
the ideal case would be to show that resistance is strongly inherited.  And we
will indeed show that, and from it will conclude that there exist election
systems that are resistant to all twenty standard types of control.

We first prove the key result, which uses the fact that $\hybrid$ can embed
its constituents to allow us to $\manyonetext$-reduce from control problems
about its constituents to control problems about $\hybrid$.

\begin{theorem}
\label{t:r-gate}
Let $k \in \nats^+$ and let $\election_0, \election_1, \ldots ,
\election_{k-1}$ be election systems.  Let $\Phi$ be one of the standard
twenty types of (constructive and destructive) control.  If for at least
one~$i$, $0 \leq i \leq k-1$, $\election_i$ is candidate-anonymous and
resistant to~$\Phi$, then $\hybrid(\election_0, \election_1, \ldots ,
\election_{k-1})$ is resistant to~$\Phi$.\footnote{
A referee presented a ``counterexample'' that, though
incorrect, is worth refuting so that no other reader need have
the same worry.
Briefly put, the
referee noticed that if $\election_0$ and $\election_1$
are candidate-anonymous election systems such that $\election_0$ is
resistant to control type $\Phi$ and $\election_1$ is not, then
$\hybrid(\election_0, \election_1)$ will be routed to $\election_1$ unless
all candidate names are even, and then said that since it will
not always be the case that all candidate names are even,
this showed that
the hybrid was not resistant.
The error in this reasoning is the incorrect implicit claim
that NP-hard problems cannot have subcases---sometimes even
quite broad
ones---in~$\p$.  They in fact can, e.g., 2CNF-SAT is in~$\p$.
}
\end{theorem}

\begin{proofs}
  Let $\Phi$ and $i$ and $\election_0, \election_1, \ldots , \election_{k-1}$
  satisfy the conditions of the theorem.  By Theorem~\ref{t:ABK},
  $\hybrid(\election_0, \election_1, \ldots , \election_{k-1})$ is susceptible
  to~$\Phi$, since $\election_i$'s resistance ensures that $\election_i$ is
  susceptible.  And since $\election_i$ is resistant to~$\Phi$, the control
  problem for $\election_i$ is $\np$-hard.  Now, note that the control problem
  for $\election_i$ with respect to $\Phi$ itself many-one polynomial-time
  reduces to the control problem for $\hybrid(\election_0, \election_1, \ldots
  , \election_{k-1})$ with respect to $\Phi$, via the reduction that replaces
  each candidate name $c$ with the name $kc+i$.  So the control problem for
  $\hybrid(\election_0, \election_1, \ldots , \election_{k-1})$ with respect
  to $\Phi$ is itself $\np$-hard.~\end{proofs}

We note in passing that there is nothing particularly limiting the above
theorem to the standard twenty types of control.  The proof's many-one
reduction would work equally well on almost any natural control type one might
imagine.

Theorem~\ref{t:r-gate} immediately yields the following corollary.

\begin{corollary}
\label{c:r-gate}
$\hybrid$ strongly inherits resistance to each of the standard twenty
types of
control.
\end{corollary}

Before we turn to applying this corollary, let us note that
Theorem~\ref{t:r-gate} and Corollary~\ref{c:r-gate} are both, as is this
entire paper, within the most natural, most typical model: Votes are input as
a list (``nonsuccinct'' input) and each vote counts equally (``unweighted''
votes).  We mention that for each of the other three cases---``succinct,
weighted,'' ``succinct, unweighted,'' and ``nonsuccinct,
weighted''---Theorem~\ref{t:r-gate} and Corollary~\ref{c:r-gate} both still
hold (by the same proofs as above).

Let us apply Corollary~\ref{c:r-gate} to obtain election systems that
are broadly resistant to control.  (We mention that since this paper
first appeared, \cite{fal-hem-hem-rot:c:copeland-alpha} (see also
\cite{fal-hem-hem-rot:c:llull}) has created a very attractive alternate
way of seeing that systems exist that resist the ten standard types of
constructive control, namely, by proving that Copeland's election
system resists the ten standard constructive control types.
Unfortunately, the just-cited papers also show that Copeland elections
do not resist some of the ten standard destructive control types.)

\begin{corollary}
\label{c:c-punch}
There exist election systems---for example,
$\hybrid(\plurality,\condorcet)$---that are resistant to all the standard ten
types of constructive control.
\end{corollary}

\begin{proofs}
  By the work of Bartholdi, Tovey, and Trick (\cite{bar-tov-tri:j:control},
  see also~\cite{hem-hem-rot:j:destructive-control} regarding the tie-handling
  rules), for each of the standard
  ten types of constructive control, at least one of $\plurality$ and
  $\condorcet$ is resistant.  So by Corollary~\ref{c:r-gate} and the
  candidate-anonymity of $\plurality$ and $\condorcet$, we are 
  done.~\end{proofs}

To make the same claim for destructive control, 
we first
construct an (artificial) system, $\electionnotallone$ (we define this
system shortly), having the missing three resistance                 
properties.  (As mentioned earlier,
an alternate approach would be to rely on work done after 
this paper, stating that the system ``Copeland${}^{0.5}$'' has these 
three resistances~\cite{fal-hem-hem-rot:c:copeland-alpha}, and then 
to use that system as a component and hybridize using it.)

\begin{lemma}
\label{l:missing-three}
There exists a candidate-anonymous election system, 
$\electionnotallone$,
that is resistant to 
\begin{enumerate}
\item\label{l:missing-three:was-part-a}
destructive control by deleting voters,
\item\label{l:missing-three:was-part-b}
destructive control by adding voters, and
\item\label{l:missing-three:was-part-c}
destructive control by partition of voters in the TE model.
\end{enumerate}
\end{lemma}

\sproof
Our proof constructs reductions from Exact Cover by Three-Sets
(X3C, for short), which is one of the standard $\np$-complete problems (see
Garey and Johnson~\cite{gar-joh:b:int}) and is defined as
follows:

\subsubsection*{Exact Cover by Three-Sets (X3C)}

\begin{description}
\item[Given:]
A set $B$, 
and a family $\mathcal{S} = \{S_1, S_2, \ldots ,
  S_n\}$ of subsets $S_i$ of $B$ with $\|S_i\| = 3$ for each~$i$.
\item[Question:] Does $\mathcal{S}$ have an exact cover for~$B$?  That is, is
  there a subfamily $\mathcal{S}' \subseteq \mathcal{S}$ such that every
  element of $B$ occurs in exactly one set in~$\mathcal{S}'$?
\end{description}

  The system $\electionnotallone$ works as follows:
  The first four candidates in each vote get one point
  each.  A candidate $c$ is a winner if and only if $c$
  is the most preferred candidate of a strict majority of the
voters and it is not the case that
  all candidates other than $c$ have one point.
  Note that if $c$ is a winner, then $c$ is the unique winner.
It is immediate that $\electionnotallone$ is candidate-anonymous
and it is easy to see that $\electionnotallone$ is 
susceptible to destructive control by
deleting voters, adding voters, and partition of voters in the TE model.

  $\np$-hardness of the
  destructive control problem follows in all three cases from a reduction from
  the problem X3C.  Let $\pair{B,\mathcal{S}}$ be an instance of X3C with
  $\|B\| = m$, $\mathcal{S} = \{S_1, S_2, \ldots , S_n\}$,
  $S_i \subseteq B$, and $\|S_i\| = 3$ for each~$i$.  Without
  loss of generality, we assume that $m$ is divisible by~$3$
and $m/3 \geq 1$.
Construct an instance of Destructive Control by Deleting Voters as
follows. The candidate set is defined as $C = B \cup \{d\}$, where
$d$ is the distinguished candidate. 
The set $V$ of voters consists of $n$ preference orders
corresponding to the sets in~$\mathcal{S}$: For each $i$,
$V$ contains voter $d > S_i > \cdots$, where ``$S_i$'' denotes the elements 
of $S_i$ in some arbitrary order,
and  ``$\cdots$'' means that the
remaining candidates follow in some arbitrary order.

\begin{claim}
\label{cla:missing-three-deleting-voters}
$\pair{B,\mathcal{S}}$ is in X3C if and only if there is a 
set $V' \subseteq V$ such that $\|V'\| \leq n - m/3$ and
$d$ is not the unique winner of $\parpair{C,V-V'}$.
\end{claim}

\sproofof{Claim~\ref{cla:missing-three-deleting-voters}} 
Suppose that $\pair{B,\mathcal{S}}$ is in X3C.  Let $V''$ consist
of $m/3$ voters from $V$ that correspond to an exact cover and
let $V' = V - V''$. Then $\|V'\| = n - m/3$ and
each candidate $c \in C$, $c \neq d$, scores one point in $V'' = V - V'$.
So, $d$ is not the unique winner of $\pair{C,V- V'}$.

For the converse, let $V' \subseteq V$ be such that $\|V'\| \leq n - m/3$ and
$d$ is not the unique winner of $\parpair{C,V-V'}$.
Since $d$ is the most preferred candidate for every voter
in $V$ and $V-V'$ is not empty (since $m/3 \geq 1$),
it follows that all other candidates
score one point in $V-V'$.
This implies that the voters in $V-V'$ correspond to
an exact cover of 
$\mathcal{S}$.~\eproofof{Claim~\ref{cla:missing-three-deleting-voters}}

Claim~\ref{cla:missing-three-deleting-voters} establishes 
Case~\ref{l:missing-three:was-part-a} of
Lemma~\ref{l:missing-three}.

We will use a similar construction to reduce X3C to Destructive Control
by Adding Voters. 
The candidate set is now defined as $C = B \cup \{d, c_1, c_2, c_3\}$.
The set $V$ of registered voters consists of one voter
$d > c_1 > c_2 > c_3 > \cdots$ and the set $W$ of
unregistered voters consists of $n$ preference orders
corresponding to the sets in~$\mathcal{S}$: For each $i$,
$W$ contains voter $d > S_i > \cdots$, where ``$S_i$'' denotes the elements 
of $S_i$ in some arbitrary order, and  ``$\cdots$'' means that the
remaining candidates follow in some arbitrary order.

\begin{claim}
\label{cla:missing-three-adding-voters}
$\pair{B,\mathcal{S}}$ is in X3C if and only if
there is a set $W' \subseteq W$ such that $\|W'\| \leq m/3$ 
and $d$ is not the unique winner of $\parpair{C,V\cup W'}$.
\end{claim}

\sproofof{Claim~\ref{cla:missing-three-adding-voters}} 
Suppose that $\pair{B,\mathcal{S}}$ is in X3C.  Let $W'$ consist
of $m/3$ voters from $W$ that correspond to an exact cover.
Then each candidate $c \in C$, $c \neq d$, scores one point in $V \cup W'$.
So, $d$ is not the unique winner of $\pair{C,V \cup W'}$.

For the converse, let $W' \subseteq W$ be such that $\|W'\| \leq m/3$ and
$d$ is not the unique winner of $\parpair{C,V\cup W'}$.
Since $d$ is the most preferred candidate for every voter
in $V \cup W'$ and $V \cup W'$ is not empty, it follows that all
other candidates score one point in $V \cup W'$.
This implies that the voters in $W'$ correspond to
an exact cover of 
$\mathcal{S}$.~\eproofof{Claim~\ref{cla:missing-three-adding-voters}}

Claim~\ref{cla:missing-three-adding-voters} establishes 
Case~\ref{l:missing-three:was-part-b} of
Lemma~\ref{l:missing-three}.

We will use a similar construction to reduce X3C to Destructive Control
by Partition of Voters in the TE model.  As in the previous case, the
candidate set is defined as $C = B \cup \{d, c_1, c_2, c_3\}$.
The set $V$ of voters consists of one voter of the form
$d > c_1 > c_2 > c_3 > \cdots$, $n$ voters of the form
$c_1 > \cdots$, and $n$ preference orders corresponding to the sets
in~$\mathcal{S}$: For each $i$,
$V$ contains voter $d > S_i > \cdots$, where ``$S_i$'' denotes the elements 
of $S_i$ in some arbitrary order, and  ``$\cdots$'' means that the
remaining candidates follow in some arbitrary order.

\begin{claim}
\label{cla:missing-three-partition-of-voters}
$\pair{B,\mathcal{S}}$ is in X3C if and only if
$d$ can be made to be not the unique winner
of this election by partition of voters in the TE model.\footnote{In fact, the
  tie-handling rule is irrelevant here, since there is always at most one
  winner in an $\electionnotallone$ election.  However, we state the claim
  for the TE model because that is what we need for
  Lemma~\ref{l:missing-three}.}
\end{claim}

\sproofof{Claim~\ref{cla:missing-three-partition-of-voters}}
Suppose that $\pair{B,\mathcal{S}}$ is in X3C. Let the 
first subelection consist of voter $d > c_1 > c_2 > c_3 > \cdots$
and $m/3$ voters of the form $d > S_i > \cdots$
that correspond to an exact cover. Then $d$ does
not win the first subelection.
$d$ also does not win the second
subelection, since most voters in the second subelection prefer
$c_1$. Since $d$ wins neither subelection,
$d$ is not an overall winner of the election.

For the converse, suppose that $d$ is not the unique winner of the
election.   Since $d$ is the unique winner of every subelection
that involves $d$ and at most one other candidate, it follows
that $d$ is the unique winner of the final run-off if $d$ participates
in the final run-off.  Since $d$ is not the unique winner of the election,
$d$ cannot participate in the final run-off, and should thus
lose both subelections.
In at least one of the subelections, $d$ is the most
preferred candidate for a strict majority of the voters.
Since $d$ does not win this subelection, the voters in this subelection of
the form $d > S_i > \cdots$ correspond to an exact 
cover.~\eproofof{Claim~\ref{cla:missing-three-partition-of-voters}}

Claim~\ref{cla:missing-three-partition-of-voters} establishes 
Case~\ref{l:missing-three:was-part-c} of
Lemma~\ref{l:missing-three}.~\eproofof{Lemma~\ref{l:missing-three}}

\begin{corollary}
\label{c:d-punch}
There exist election systems that are resistant to all ten standard types of
destructive control.
\end{corollary}

\begin{proofs}
  By Lemma~\ref{l:missing-three} and the work of Hemaspaandra, Hemaspaandra,
  and Rothe \cite{hem-hem-rot:j:destructive-control}, for each of the
  standard ten types of destructive control, at least one of $\plurality$ and
  $\electionnotallone$ is resistant.  So by Corollary~\ref{c:r-gate} and the
  candidate-anonymity of $\plurality$ and $\electionnotallone$, we are 
  done.~\end{proofs}

We cannot apply Theorem~\ref{t:r-gate} directly to rehybridize the
systems of Corollaries~\ref{c:c-punch} and~\ref{c:d-punch}, because
$\hybrid$ itself is not in general candidate-anonymous.  However, we
can get the same conclusion by directly hybridizing all the
constituents underlying Corollaries~\ref{c:c-punch}
and~\ref{c:d-punch}.  
\begin{theorem}
\label{t:all-twenty}
There exist election systems that are resistant to all twenty standard
 types of control.
\end{theorem}
The proof simply is to consider
$\hybrid(\plurality,\condorcet,\electionnotallone)$.

\subsection{Immunity}

We now turn to inheritance of immunity.  Here, for each of
constructive and destructive control, five cases inherit and five
cases provably fail to inherit.

However, let is first pause a moment 
to define the artificial, candidate-anonymous
election system $\electionnull$, in which all
candidates always lose.  Note that $\electionnull$
is immune
to all twenty standard types of
control---trivially, since no one ever wins.  This type of result shows
why it is natural to focus, as this paper does, on achieving broad
resistance rather than on achieving broad immunity: Resistance is less
trivializable than immunity.  In fact, we will now turn to
showing that immunity to constructive control by deleting candidates is
inherited by $\hybrid$
(Theorem~\ref{t:hybrid-inherits-immunity-to-constr-control-by-deleting-candidates}),
and our proof of that
will be based on a trivialization of immunity.  In that proof,
we will employ the following somewhat interesting triviality result.

\begin{theorem}
\label{t:special}
Any candidate-anonymous election system that is immune to
constructive
control by deleting candidates can never have a unique winner.
\end{theorem}

\begin{proofs}
Let $\election$ be any candidate-anonymous election system that is
immune to constructive control by deleting candidates.  Suppose there
is an election $\parpair{C,V}$ 
in which some candidate, say $c \in C$, is the unique winner, and let $V$
consist of the voters $v_1, v_2, \ldots , v_k$.  By
candidate-anonymity of~$\election$, there is a candidate set $D$ with
$D \cap C = \emptyset$ that simply renames all candidates in~$C$, and
a voter set $W$ consisting of voters $w_1, w_2, \ldots , w_k$
that simply renames all
candidates occurring in~$V$ from their $C$ names to their $D$ names
such that $d \in D$ is the unique winner of $\parpair{D,W}$.  Now, consider
the election with candidate set $C \cup D$ and voter set $X$
consisting of the $k$ preference orders
\[
v_{i,1} > v_{i,2} > \cdots >  v_{i,\|C\|} > 
w_{i,1} > w_{i,2} > \cdots >  w_{i,\|D\|} ,
\]
where $1 \leq i \leq k$ and the $i$th voter in, respectively, $V$ and
$W$ has preference order $v_i = v_{i,1} > v_{i,2} > \cdots > 
v_{i,\|C\|}$ and $w_i = w_{i,1} > w_{i,2} > \cdots >  w_{i,\|D\|}$.

If $c$ is not a unique winner of the election $\parpair{C \cup D,X}$, then
$\election$ is  not immune to constructive control by deleting
candidates, as by deleting $D$, $c$ becomes the
unique winner.  Similarly, if $d$ is not a unique winner of the
election $\parpair{C \cup D,X}$, then $\election$ is not immune to
constructive control by deleting candidates, as by deleting $C$, $d$
becomes the
unique winner.  However, since $\election$ is immune to constructive
control by deleting candidates and since $\parpair{C \cup D,X}$ cannot have
both $c$ and $d$ be unique winners, our supposition that $\election$
can have a unique winner is contradicted.~\end{proofs}

The following lemma follows immediately from the definition of $\hybrid$.

\begin{lemma}
\label{l:never}
``Never having
a unique winner'' is inherited by $\hybrid$.
\end{lemma}

We now can prove our claim.

\begin{theorem}
\label{t:hybrid-inherits-immunity-to-constr-control-by-deleting-candidates}
Immunity to constructive control by deleting candidates is inherited by
$\hybrid$.
\end{theorem}

\begin{proofs}
If $\election_0, \election_1, \ldots , \election_{k-1}$ are
candidate-anonymous election systems that are immune to constructive
control by deleting candidates, then by Theorem~\ref{t:special} each
never has a unique winner.  So by Lemma~\ref{l:never},
$\hybrid(\election_0, \election_1, \ldots , \election_{k-1})$ never
has a unique winner. 
This proves
that $\hybrid(\election_0, \election_1, \ldots , \election_{k-1})$ is
immune to every type of control.~\end{proofs}

By applying a duality result of Hemaspaandra, Hemaspaandra, and
Rothe $k$ times, we can retarget this to a type of destructive
control.

\begin{proposition}[\cite{hem-hem-rot:j:destructive-control}]~
\label{p:duality}
\begin{enumerate}
\item
\label{p:duality-part-a}
A voting system is 
susceptible to constructive control
by adding candidates if and only if it is 
susceptible to destructive control
by deleting candidates.

\item
\label{p:duality-part-b}
A voting system is 
susceptible to constructive control
by deleting candidates if and only if it is 
susceptible to destructive control
by adding candidates.
\end{enumerate}
\end{proposition}

\begin{corollary}
\label{c:hybrid-inherits-immunity-to-destr-control-by-adding-candidates}
Immunity to destructive control by adding candidates is inherited by
$\hybrid$.
\end{corollary}

\begin{proofs}
The proof follows immediately from
Theorem~\ref{t:hybrid-inherits-immunity-to-constr-control-by-deleting-candidates}
plus $1 + (k-1)$ applications of
Proposition~\ref{p:duality}.\ref{p:duality-part-b}.~\end{proofs}

$\hybrid$'s immunity to all voter-related types of control is immediate.

\begin{theorem}
\label{t:eight-hybrid-inherits-immunity-results}
Immunity to constructive and destructive control under each of these:
\begin{enumerate}
\item\label{t:eight-hybrid-inherits-immunity-results:was-part-a}
adding voters,
\item\label{t:eight-hybrid-inherits-immunity-results:was-part-b}
deleting voters,
\item\label{t:eight-hybrid-inherits-immunity-results:was-part-c}
partition of voters in model~TE, and
\item\label{t:eight-hybrid-inherits-immunity-results:was-part-d}
partition of voters in model~TP
\end{enumerate}
is inherited (and flexibly so) by $\hybrid$.
\end{theorem}

\begin{proofs}
Since none of this theorem's four operations
changes the candidate set, $\hybrid$ routes all
election evaluations related to a given input to the same election
system~$\election_i$, and so by that system's immunity the theorem 
holds.~\end{proofs}

For the ten remaining cases, inheritance does not hold.

\begin{theorem}
\label{t:immunity-not-inherited-results}
$\hybrid$ does not inherit immunity to:
\begin{enumerate}
\item\label{t:immunity-not-inherited-results:was-part-a}
constructive control by adding candidates,
\item\label{t:immunity-not-inherited-results:was-part-b}
destructive control by deleting candidates,
\item\label{t:immunity-not-inherited-results:was-part-c}
constructive control by partition and run-off
partition of candidates (both in model~TE and~TP), and
\item\label{t:immunity-not-inherited-results:was-part-d}
destructive control by partition and run-off partition
of candidates (both in model~TE and~TP).
\end{enumerate}
\end{theorem}

\begin{proofs} For each of these ten control problems, we provide a
counterexample showing that immunity is not inherited by $\hybrid$.
We will use the artificial, candidate-anonymous
election systems $\electiononefirst$ and $\electionlast$, which
are defined as follows.
A candidate $c$ wins an $\electiononefirst$ election if
and only if there is one voter
and this voter ranks $c$ first.  
A candidate $c$ wins an $\electionlast$ election if
and only if there is one voter
and this voter ranks $c$ last.  It is immediate that
$\electiononefirst$ and $\electionlast$ are candidate-anonymous.

\begin{enumerate}

\item\label{proof-of-t:immunity-not-inherited-results:was-part-a}
It is immediate that $\electiononefirst$ and $\electionlast$
are both immune to constructive control by adding candidates.
However, $\hybrid(\electiononefirst,\electionlast)$
is susceptible to constructive control by adding candidates.
To see that, consider the election with set $C = \{0,2\}$ of
qualified candidates, set $D = \{1\}$ of possible spoiler
candidates, and voter set $V$ consisting of one voter with
preference order $2 > 1 > 0$.  Candidate~$0$ loses in
$\hybrid(\electiononefirst,\electionlast)$ in
election $\parpair{C,V}$: All candidate names in $C$ are congruent to $0 \bmod 2$,
so $\hybrid$ routes its input to $\electiononefirst$.
However, if the spoiler candidate~$1$ is added to $C$ and so the
election $\parpair{C \cup D,V}$ is input to
$\hybrid(\electiononefirst,\electionlast)$, this switches $\hybrid$
to now implement $\electionlast$, and so candidate~$0$ now
is the unique winner of this election.
Thus, $\hybrid(\electiononefirst,\electionlast)$ is not
immune to constructive control by adding candidates.

\item\label{proof-of-t:immunity-not-inherited-results:was-part-b}
This claim follows immediately from 
Part~\ref{t:immunity-not-inherited-results:was-part-a} and 
Proposition~\ref{p:duality}.\ref{p:duality-part-a}.

\item\label{proof-of-t:immunity-not-inherited-results:was-part-c}
It is easy to see that $\electiononefirst$ and $\electionlast$
are immune to constructive control by partition and run-off
partition of candidates in models TE and TP: No matter how we partition
the candidates, there can only be winners if there is exactly one voter.
And in an election with one voter, the first-ranked (respectively,
last-ranked)
candidate is the unique winner of the election for all partitions of 
candidates in all partitioning models.  

However, $\hybrid(\electiononefirst,\electionlast)$ is susceptible to
constructive control by partition and run-off partition
of candidates in models TE
and~{TP}.  To see that, consider the election with candidate set $C =
\{0,1,2\}$, distinguished candidate~$2$, and voter set $V$
consisting of one voter with preference order $2 > 1 > 0$.
On input $\parpair{C,V}$, $\hybrid(\electiononefirst,\electionlast)$ routes the
election to $\electionlast$, so $0$ is the unique winner of the election.
Now consider the candidate partition $C_1 = \{0,1\}$ and $C_2 = \{2\}$.
Then 0 is the unique winner of $\parpair{C_1,V}$ and 
2 is the unique winner of $\parpair{C_2,V}$.  So, in both partition
and run-off partition of candidates in models TE and TP,
the final round of the election is $\parpair{\{0,2\},V}$.
Since all candidate names are congruent to $0 \bmod 2$,
$\hybrid$ routes 
its input to $\electiononefirst$, so $2$ is the unique winner.
Since 2 has been turned from a nonwinner into
a unique winner, it follows that $\hybrid(\electiononefirst,\electionlast)$
is susceptible to constructive control by partition and run-off partition
of candidates in models~TE and~{TP}.

\item\label{proof-of-t:immunity-not-inherited-results:was-part-d}
The argument given at the start of 
Part~\ref{proof-of-t:immunity-not-inherited-results:was-part-c}
of this proof also shows that
$\electiononefirst$ and $\electionlast$
are immune to destructive control by partition and run-off partition
of candidates in models~TE and~TP.
To show that $\hybrid(\electiononefirst,\electionlast)$ is susceptible to
destructive control by partition and run-off partition of candidates
in models TE and TP,
it suffices to note that the candidate partition of 
Part~\ref{proof-of-t:immunity-not-inherited-results:was-part-c}
of this proof turns
candidate $0$ from a unique winner into a nonwinner.\nopagebreak%
\end{enumerate}\nopagebreak\end{proofs}

\subsection{Vulnerability}\label{ss:vulnerable}

$\hybrid$ strongly inherited resistance, which is precisely what one
wants, since that is both the aesthetically
pleasing case and broadens resistance to control.  However,
for vulnerability it is less clear what outcome to root for.
Inheritance would be the mathematically more beautiful outcome.  But
on the other hand, what inheritance would inherit is vulnerability,
and vulnerability to control is in general a bad thing---so maybe one
should hope for ``Not I(nherits)'' entries
for our tables in this
column.  In fact, our results here are mixed.
In particular, we for ten cases prove that inheritance
holds unconditionally and for ten cases prove that inheritance holds
(though in some cases we have to limit ourselves to election systems
with winner/unique winner problems that fall into the polynomial hierarchy)
if and only if $\p = \np$.
Since any practically useful election system should have 
winner/unique winner
problems in~$\p$, this is not an onerous limitation. Indeed, even such
``difficult'' election systems as those of Dodgson, Kemeny, and Young have
winner/unique winner problems in the polynomial hierarchy.

We start with two vulnerability results that can be established by direct
algorithmic attacks.

\begin{theorem}
\label{t:vulnerability-inherited-for-adding-candidates}
$\hybrid$ inherits vulnerability to both constructive and destructive control
by adding candidates.
\end{theorem}

\begin{proofs}
  We start with the constructive case.  Let $k \in \nats^+$ and let
  $\election_0, \election_1, \ldots , \election_{k-1}$ be candidate-anonymous
  election systems each vulnerable to constructive control by adding
  candidates.  Let $A_i$, $0 \leq i \leq k-1$, be the polynomial-time
  algorithms showing that $\election_i$ is vulnerable to constructive control
  by adding candidates.
  
  Since $\election_0$ is vulnerable to constructive control by adding
candidates, $\election_0$ is susceptible to constructive control by adding
candidates, and it
  follows from Theorem~\ref{t:ABK} that $\hybrid(\election_0,
  \election_1, \ldots , \election_{k-1})$ 
  is susceptible to constructive control by adding candidates.

  To complete our proof of vulnerability, we turn to seeking a
  polynomial-time algorithm for this control problem.  Given an instance of
  Constructive Control by Adding Candidates, let $Q$ be the set of qualified
  candidates with $c \in Q$ being the distinguished candidate, let $S$ be the
  set of spoiler candidates, and let $V$ be the voter set with preferences
  over $Q \cup S$.  We distinguish the following three cases.
\begin{case}
\item\label{t:vulnerability-inherited-for-adding-candidates:case-1}
  \emph{Not all candidates in $Q$ have the same value modulo
    $k$.}\quad Then, regardless of whether none, some, or all candidates from
  $S$ are added, the initial and final election will be routed by
  $\hybrid(\election_0, \election_1, \ldots , \election_{k-1})$ to
  $\electiondefault = \election_{k-1}$.  Thus, we can simply invoke
  algorithm $A_{k-1}$.
  
\item\label{t:vulnerability-inherited-for-adding-candidates:case-2}
   \emph{All candidates in $Q$ have the same value modulo $k$, call
    it~$q$, and all candidates in~$S$ also have values congruent to $q \bmod
    k$.}\quad Then we simply invoke algorithm $A_q$.
  
\item\label{t:vulnerability-inherited-for-adding-candidates:case-3}
   \emph{All candidates in $Q$ have the same value modulo $k$, call
    it~$q$, and at least one candidate in $S$ has a value not congruent to $q
    \bmod k$.}\quad Then we do the following:
\begin{algorithmus}
\item \label{alg:vulnerability-inherited-for-adding-candidates:step1} Invoke
  $A_q$ to see if victory (i.e., being the unique winner) for $c$ is possible
  in $\election_q$ with $Q$ as the set of qualified candidates,
  $S_q$ as the set of  
  spoiler candidates, where $S_q$ contains all members of $S$ whose
  values are congruent to $q \bmod k$, and with voter set $V$ (restricted
to the participating candidates).
  If algorithm $A_q$ says YES, then output YES and halt; otherwise go to
  Step~\ref{alg:vulnerability-inherited-for-adding-candidates:step2}.
  
\item \label{alg:vulnerability-inherited-for-adding-candidates:step2} For each
  $d \in S \cap \{i \condition i \not\equiv q \bmod k\}$,
 invoke $A_{k-1}$ to
  see if victory for $c$ is possible in $\electiondefault = \election_{k-1}$
  with $Q \cup \{d\}$ as the set of qualified candidates,
  $S' = S - \{d\}$ as the set of of spoiler candidates,
  and with voter set~$V$.
  If algorithm $A_{k-1}$ says YES, then output YES and halt;
  otherwise go to the next iteration of this for-each loop.

\item \label{alg:vulnerability-inherited-for-adding-candidates:step3} If we
  reach this step (i.e., the for-each loop in
  Step~\ref{alg:vulnerability-inherited-for-adding-candidates:step2} was not
  successful in any of its iterations), then output ``victory for $c$ is
  impossible'' and halt.
\end{algorithmus}
\end{case}

Note that the reason we in 
Case~\ref{t:vulnerability-inherited-for-adding-candidates:case-3}
had to include the
``try each
non-mod-$q$ as being forced in'' flourish is that only when at least one
non-mod-$q$ candidate is in the candidate set will we switch from
$\election_q$
to $\electiondefault$,
so we must be very careful to only
use $\electiondefault$ regarding cases that will be routed to
it.  We cannot in any obvious way replace the loop with a single
test, as a single vulnerability test does not let us address
``Is there some subset of the spoiler set that includes at least one
marked spoiler and that makes $c$ win?'' as that is a different problem,
though the algorithm in effect does that via a polynomial number of
ordinary vulnerability tests.

As to the destructive case, it follows 
from Theorem~\ref{t:ABK} that $\hybrid(\election_0,
\election_1, \ldots , \election_{k-1})$ 
is susceptible to destructive control by adding candidates.
To show that vulnerability inherits,
exactly the same algorithm as above works, except
with each time ``victory'' occurs replaced with ``not being a unique winner,''
and of course when we now call the underlying polynomial-time 
algorithms for~$\election_i$,
they will be for the case of destructive control by adding 
candidates.~\end{proofs}

We now state eight trivial inheritance cases.

\begin{theorem}
\label{t:vulnerability-inherited-for-voter-control}
Vulnerability to constructive and destructive control by adding voters,
deleting voters, partition of voters in model~TE, and partition of voters in
model~TP are all inherited
by $\hybrid$.
\end{theorem}

\begin{proofs}
  None of these operations changes the candidate set, so whatever system
  $\hybrid$ routes the input to ($\election_0$ or $\election_1$ or \ldots\ or
  $\election_{k-1}$ or $\electiondefault = \election_{k-1}$) will handle all
  the election evaluations related to the given input, and so we can directly
  use \emph{its} polynomial-time algorithm showing vulnerability
  to the given type of control.  Susceptibility follows from
  Theorem~\ref{t:ABK}.~\end{proofs}

We are down to ten cases---deleting candidates and the four
candidate-based partition cases, with each of those five cases
appearing in constructive control and destructive control
versions.  For these ten cases,
we will prove that inheritance holds
(though in some cases we have to limit ourselves to election systems
whose winner/unique winner problems are in the polynomial hierarchy)
if and only if $\p = \np$.

\begin{theorem}
\label{t:nonmess}
Let $\Phi$ be one of the following two control types:
\begin{enumerate}
\item constructive control by deleting candidates,
\item destructive control by deleting candidates,
\end{enumerate}
let $k \in \nats^+$, and let $\election_0, \election_1, \ldots ,
\election_{k-1}$ be election systems
whose $\Phi$-control problems are in the
polynomial hierarchy and such that
for at least one~$i$, $0 \leq i \leq k-1$,
$\election_i$ is candidate-anonymous and
susceptible to~$\Phi$.
If $\p = \np$ then $\hybrid(\election_0, \election_1,
\ldots , \election_{k-1})$ is vulnerable to~$\Phi$.
\end{theorem}

\begin{proofs}
It follows immediately from Theorem~\ref{t:ABK} that 
$\hybrid(\election_0, \election_1, \ldots , \election_{k-1})$ is susceptible
to~$\Phi$.  To show vulnerability, note that for constructive control
by deleting candidates
the unique winner problem $\manyonetext$-reduces (via the trick mentioned in
  Section~\ref{s:prelims-isvr}: asking
  about deleting zero candidates) to the control problem.
And for destructive control by deleting candidates,
by the same trick, we have that the unique winner problem
$\manyonetext$-reduces to the complement of the control problem.
So in both cases,
  since $\p =
  \np$ is assumed and the control problems are assumed to be in the polynomial
  hierarchy, we do have a polynomial-time
  algorithm for $\hybrid(\election_0, \election_1, \ldots , \election_{k-1})$'s
  unique winner problem by Proposition~\ref{t:winner-p}.
  Now, guess every possible way of legally deleting
  candidates and using the polynomial-time algorithm for $\hybrid(\election_0,
  \election_1, \ldots , \election_{k-1})$ see if under some such way our goal
  (making a specified candidate the unique winner or precluding a specified
  candidate from being a
  unique winner) can be achieved.  This is an $\np$ algorithm but since $\p =
  \np$ it proves vulnerability to~$\Phi$.~\end{proofs}

\begin{corollary}
\label{c:two-off}
If $\p = \np$ then $\hybrid$ inherits
vulnerability to
constructive and destructive control by deleting candidates.
\end{corollary}

On the other hand, we have the following result.  
\emph{Since the reader has earlier in this paper seen many examples 
of our resistance proofs and has just seen some vulnerability-related
constructions, for the rest of this 
section we defer the proofs to Appendix~\ref{a:defer}.}

\begin{theorem}
\label{t:one-off}
There exist candidate-anonymous election systems $\election_0$ and $\election_1$ that are
vulnerable to constructive control by deleting candidates but such that
$\hybrid(\election_0,\election_1)$ is resistant to 
constructive control by deleting candidates.
\end{theorem}

From Corollary~\ref{c:two-off}, Theorem~\ref{t:one-off}, and the fact
that if $\p \neq \np$ then resistance implies nonvulnerability,
we have our characterization.

\begin{corollary}
\label{c:characterization-constr-control-deleting-cand}
$\p = \np$ if and only if $\hybrid$ inherits vulnerability to constructive
control by deleting candidates.
\end{corollary}

The corresponding characterization for the case of destructive
control by deleting candidates will be established later
as Corollary~\ref{c:characterization-destr-control-deleting-cand}.
We
note that the proof of Theorem~\ref{t:one-off} also immediately
establishes the following, by taking $\election_0 = \electionnull$,
the election system in which all candidates always lose.

\begin{corollary}
\label{c:one-off}
There exist candidate-anonymous election systems $\election_0$ and $\election_1$ such
that $\election_0$ is immune
to constructive control by deleting candidates 
and $\election_1$ is
vulnerable to constructive control by deleting candidates but such that
$\hybrid(\election_0,\election_1)$ is resistant to 
constructive control by deleting candidates.
\end{corollary}

Unfortunately, the analogs of Theorem~\ref{t:nonmess} for candidate-related
partitions are less clean.  Regarding the strengths in the hypotheses in the
following theorem, we mention that when a winner problem is in
the polynomial hierarchy, that
clearly implies that the corresponding unique winner problem is
in the polynomial hierarchy. 

\begin{theorem}
\label{t:let}
\begin{enumerate}
\item\label{t:let:was-part-a}
Let $\Phi$ be one of the following four control types:
\begin{enumerate}
\item\label{t:let:was-part-a-1}
 constructive control by partition of candidates in model~TP,
\item\label{t:let:was-part-a-2}
 destructive control by partition of candidates in model~TP,
\item\label{t:let:was-part-a-3}
 constructive control by run-off partition of candidates in model~TP,
\item\label{t:let:was-part-a-4}
 destructive control by run-off partition of candidates in model~TP,
\end{enumerate}
let $k \in \nats^+$, and let $\election_0, \election_1, \ldots ,
\election_{k-1}$ be election systems such that
for at least one~$i$, $0 \leq i \leq k-1$,
$\election_i$ is candidate-anonymous and 
susceptible to~$\Phi$.
If $\p = \np$
and the winner problem for $\hybrid(\election_0, \election_1, \ldots ,
\election_{k-1})$ is in the polynomial hierarchy,
then $\hybrid(\election_0, \election_1,
\ldots , \election_{k-1})$ is vulnerable to~$\Phi$.

\item\label{t:let:was-part-b}
Let $\Phi$ be one of the following four control types:
\begin{enumerate}
\item\label{t:let:was-part-b-1}
 constructive control by partition of candidates in model~TE,
\item\label{t:let:was-part-b-2}
 destructive control by partition of candidates in model~TE,
\item\label{t:let:was-part-b-3}
 constructive control by run-off partition of candidates in model~TE,
\item\label{t:let:was-part-b-4}
 destructive control by run-off partition of candidates in model~TE,
\end{enumerate}
let $k \in \nats^+$, and let $\election_0, \election_1, \ldots ,
\election_{k-1}$ be election systems such that
for at least one~$i$, $0 \leq i \leq k-1$,
$\election_i$ is candidate-anonymous and 
susceptible to~$\Phi$.
If $\p = \np$
and the unique winner problem for $\hybrid(\election_0, \election_1, \ldots ,
\election_{k-1})$ is in the polynomial hierarchy,
then $\hybrid(\election_0, \election_1,
\ldots , \election_{k-1})$ is vulnerable to~$\Phi$.
\end{enumerate}
\end{theorem}

In light of Proposition~\ref{t:winner-p},
Theorem~\ref{t:let} certainly implies
the following weaker but more naturally formed result.

\begin{corollary}
\label{c:simplified}
\begin{enumerate}
\item\label{c:simplified:was-part-a}
 Let $\Phi$ be one of the following four control types:
\begin{enumerate}
\item constructive control by partition of candidates in model~TP,
\item destructive control by partition of candidates in model~TP,
\item constructive control by run-off partition of candidates in model~TP,
\item destructive control by run-off partition of candidates in model~TP,
\end{enumerate}
let $k \in \nats^+$, and let $\election_0, \election_1, \ldots ,
\election_{k-1}$ be election systems such that
for at least one~$i$, $0 \leq i \leq k-1$,
$\election_i$ is candidate-anonymous and 
susceptible to~$\Phi$.
If $\p = \np$
and the winner problems for each of $\election_0, \election_1, \ldots ,
\election_{k-1}$ are in the polynomial hierarchy,
then $\hybrid(\election_0, \election_1,
\ldots , \election_{k-1})$ is vulnerable to~$\Phi$.

\item\label{c:simplified:was-part-b}
 Let $\Phi$ be one of the following four control types:
\begin{enumerate}
\item constructive control by partition of candidates in model~TE,
\item destructive control by partition of candidates in model~TE,
\item constructive control by run-off partition of candidates in model~TE,
\item destructive control by run-off partition of candidates in model~TE,
\end{enumerate}
let $k \in \nats^+$, and let $\election_0, \election_1, \ldots ,
\election_{k-1}$ be election systems such that
for at least one~$i$, $0 \leq i \leq k-1$,
$\election_i$ is candidate-anonymous and 
susceptible to~$\Phi$.
If $\p = \np$
and the unique winner problems for each of $\election_0, \election_1, \ldots ,
\election_{k-1}$ are in the polynomial hierarchy,
then $\hybrid(\election_0, \election_1,
\ldots , \election_{k-1})$ is vulnerable to~$\Phi$.
\end{enumerate}
\end{corollary}

On the other hand, we have the following result.

\begin{theorem}
\label{t:edith-147.A}
There exist candidate-anonymous election systems $\election_0$ and $\election_1$ that are
vulnerable to constructive control by run-off partition of candidates 
(in models TE and~TP) but such that 
$\hybrid(\election_0,\election_1)$ is resistant to
constructive control by run-off partition of candidates 
(in models TE and~TP).
\end{theorem}

\begin{corollary}
\label{c:edith-147.A}
If $\p \neq \np$ then $\hybrid$ does not inherit vulnerability to
constructive control by run-off partition of candidates in model~TE or~TP.
\end{corollary}

Corollaries~\ref{c:simplified} and~\ref{c:edith-147.A}
certainly imply the following result. 

\begin{corollary}
\label{c:characterization-constr-control-run-off-part-cand}
\begin{enumerate}
\item The following are equivalent.
\begin{enumerate}
\item $\hybrid$ inherits vulnerability to
constructive control by run-off partition of candidates in model TE
on election systems having unique winner problems in the polynomial hierarchy.
\item $\hybrid$ strongly inherits vulnerability to
constructive control by run-off partition of candidates in model TE
on election systems having unique winner problems in the polynomial hierarchy.
\item $\p = \np$.
\end{enumerate}

\item
The following are equivalent.
\begin{enumerate}
\item $\hybrid$ inherits vulnerability to
constructive control by run-off partition of candidates in model TP
on election systems having winner problems in the polynomial hierarchy.
\item $\hybrid$ strongly inherits vulnerability to
constructive control by run-off partition of candidates in model TP
on election systems having winner problems in the polynomial hierarchy.
\item $\p = \np$.
\end{enumerate}
\end{enumerate}
\end{corollary}

We
note that the proof 
(in Appendix~\ref{a:defer}) 
of Theorem~\ref{t:edith-147.A} 
also immediately
establishes the following, by taking $\election_0 = \electionnull$,
the election system in which all candidates always lose.

\begin{corollary}
\label{c:edith-147.A-immune}
There exist candidate-anonymous election systems $\election_0$ and $\election_1$
such
that $\election_0$ is immune
to constructive control by run-off partition of candidates
(in models~TE and~TP) 
and $\election_1$ is
vulnerable to constructive control by run-off partition of candidates
(in models~TE and~TP) 
but such that $\hybrid(\election_0,\election_1)$ is resistant to 
constructive control by run-off partition of candidates
(in models~TE and~TP).
\end{corollary}

We now turn to the analogs of Theorem~\ref{t:edith-147.A} and
Corollaries~\ref{c:edith-147.A},
\ref{c:characterization-constr-control-run-off-part-cand},
and~\ref{c:edith-147.A-immune} for 
constructive control by partition of candidates.

\begin{theorem}
\label{t:partition}
There exist candidate-anonymous election systems $\election_0$ and $\election_1$ that are
vulnerable to constructive control by partition of candidates 
(in models TE and~TP) but such that 
$\hybrid(\election_0,\election_1)$ is resistant to
constructive control by partition of candidates 
(in models TE and~TP).
\end{theorem}

\begin{corollary}
\label{c:partition}
If $\p \neq \np$ then $\hybrid$ does not inherit vulnerability to
constructive control by partition of candidates in model~TE or~TP.
\end{corollary}

Corollaries~\ref{c:simplified} and~\ref{c:partition}
certainly imply the following result. 

\begin{corollary}
\label{c:characterization-constr-control-partition-cand}
\begin{enumerate}
\item The following are equivalent.
\begin{enumerate}
\item $\hybrid$ inherits vulnerability to
constructive control by partition of candidates in model TE
on election systems having unique winner problems in the polynomial hierarchy.
\item $\hybrid$ strongly inherits vulnerability to
constructive control by partition of candidates in model TE
on election systems having unique winner problems in the polynomial hierarchy.
\item $\p = \np$.
\end{enumerate}

\item
The following are equivalent.
\begin{enumerate}
\item $\hybrid$ inherits vulnerability to
constructive control by partition of candidates in model TP
on election systems having winner problems in the polynomial hierarchy.
\item $\hybrid$ strongly inherits vulnerability to
constructive control by partition of candidates in model TP
on election systems having winner problems in the polynomial hierarchy.
\item $\p = \np$.
\end{enumerate}
\end{enumerate}
\end{corollary}

We
note that the proof 
(in Appendix~\ref{a:defer}) 
of Theorem~\ref{t:partition} 
also immediately
establishes the following, by taking $\election_0 = \electionnull$,
the election system in which all candidates always lose.

\begin{corollary}
\label{c:edith-partition-immune}
There exist candidate-anonymous election systems $\election_0$ and $\election_1$
such
that $\election_0$ is immune
to constructive control by partition of candidates
(in models~TE and~TP) 
and $\election_1$ is
vulnerable to constructive control by  partition of candidates
(in models~TE and~TP) 
but such that $\hybrid(\election_0,\election_1)$ is resistant to 
constructive control by  partition of candidates
(in models~TE and~TP). 
\end{corollary}

We now turn to the destructive control cases.

\begin{theorem}
\label{t:destructive-vulnerability-not-inherits}
There exist candidate-anonymous election systems
$\election_0$ and $\election_1$ that are
vulnerable to destructive control by deleting candidates,
partition of candidates (in models TE and~TP), and
run-off partition of candidates (in models TE and~TP), 
but such that 
$\hybrid(\election_0,\election_1)$ is resistant to
destructive control by deleting candidates,
partition of candidates (in models TE and~TP), and
run-off partition of candidates (in models TE and~TP).
\end{theorem}

From Corollary~\ref{c:two-off},
Theorem~\ref{t:destructive-vulnerability-not-inherits}
and the fact
that if $\p \neq \np$ then resistance implies nonvulnerability,
we obtain a complete characterization of the destructive control
by deleting candidates case.

\begin{corollary}
\label{c:characterization-destr-control-deleting-cand}
$\p = \np$ if and only if $\hybrid$ inherits vulnerability to destructive
control by deleting candidates.                             
\end{corollary}

Recall that we don't quite have the analog of Corollary~\ref{c:two-off}
for the partition cases. Theorem~\ref{t:destructive-vulnerability-not-inherits}
established the following implication. 

\begin{corollary}
\label{c:destructive-vulnerability-not-inherits}
If $\p \neq \np$ then $\hybrid$ does not inherit vulnerability to
destructive control by partition of candidates (in models TE and~TP) and
run-off partition of candidates (in models TE and~TP).
\end{corollary}

Corollaries~\ref{c:simplified}
and~\ref{c:destructive-vulnerability-not-inherits}
certainly
imply the following result.

\begin{corollary}
\label{c:characterization-destr-control-partition-cand}
\begin{enumerate}
\item The following are equivalent.
\begin{enumerate}
\item $\hybrid$ inherits vulnerability to
destructive control by partition of candidates in model TE
on election systems having unique winner problems in the polynomial hierarchy.
\item $\hybrid$ strongly inherits vulnerability to
destructive control by partition of candidates in model TE
on election systems having unique winner problems in the polynomial hierarchy.
\item $\p = \np$.
\end{enumerate}

\item
The following are equivalent.
\begin{enumerate}
\item $\hybrid$ inherits vulnerability to
destructive control by partition of candidates in model TP
on election systems having winner problems in the polynomial hierarchy.
\item $\hybrid$ strongly inherits vulnerability to
destructive control by partition of candidates in model TP
on election systems having winner problems in the polynomial hierarchy.
\item $\p = \np$.
\end{enumerate}

\item The following are equivalent.
\begin{enumerate}
\item $\hybrid$ inherits vulnerability to
destructive control by run-off partition of candidates in model TE
on election systems having unique winner problems in the polynomial hierarchy.
\item $\hybrid$ strongly inherits vulnerability to
destructive control by run-off partition of candidates in model TE
on election systems having unique winner problems in the polynomial hierarchy.
\item $\p = \np$.
\end{enumerate}

\item
The following are equivalent.
\begin{enumerate}
\item $\hybrid$ inherits vulnerability to
destructive control by run-off partition of candidates in model TP
on election systems having winner problems in the polynomial hierarchy.
\item $\hybrid$ strongly inherits vulnerability to
destructive control by run-off partition of candidates in model TP
on election systems having winner problems in the polynomial hierarchy.
\item $\p = \np$.
\end{enumerate}
\end{enumerate}
\end{corollary}

We
note that the proof 
(in Appendix~\ref{a:defer}) 
of
Theorem~\ref{t:destructive-vulnerability-not-inherits}
also immediately
establishes the following, by taking $\election_0$ to be the
election system where $c$ wins if and only if
$c$ is first in the first voter.

\begin{corollary}
\label{c:destructive-vulnerability-immunity-not-inherits}
There exist candidate-anonymous election systems $\election_0$ and $\election_1$
such
that $\election_0$ is immune
to destructive control by deleting candidates,
partition of candidates (in models TE and~TP), and
run-off partition of candidates (in models TE and~TP), 
(in models~TE and~TP) 
and $\election_1$ is vulnerable to
to destructive control by deleting candidates,
partition of candidates (in models TE and~TP), and
run-off partition of candidates (in models TE and~TP), 
but such that $\hybrid(\election_0,\election_1)$ is resistant to 
destructive control by deleting candidates,
partition of candidates (in models TE and~TP), and
run-off partition of candidates (in models TE and~TP).
\end{corollary}

\section{Conclusions}

Tables~\ref{tab:constructive} and~\ref{tab:destructive} summarize our
inheritance results.  The main contribution of this paper is the $\hybrid$
system, the fact that $\hybrid$ strongly inherits resistance, and the
consequence 
that there is an election system that resists all twenty
standard types of electoral control.
A very natural direction for future research is to broadly
explore the control behavior of standard election systems, and 
in particular to see if any 
exhibit broad resistance
to control.

\medskip

\noindent
\paragraph*{ Acknowledgments:}
We thank 
the 
COMSOC-06 and IJCAI-07
meetings for including our work, and 
Holger Spakowski and 
anonymous referees 
for helpful comments.

\bibliographystyle{unsrt}
%
%\bibliography{gry}

\appendix
\section{Deferred Proofs from Section~\protect\ref{ss:vulnerable}}
\label{a:defer}

We now give the proofs deferred from Section~\ref{ss:vulnerable}.

\medskip

\sproofof{Theorem~\ref{t:one-off}}
  In~$\election_0$, each candidate wins if and only if it is the only
  candidate.  Clearly, $\election_0$ is candidate-anonymous and
  susceptible to constructive control by
  deleting candidates, and it is a polynomial-time task to decide when control
  can be asserted, so $\election_0$ is vulnerable to this type of control.
  
  Election system $\election_1$ is defined as follows.
  For each candidate $c \in C$, $c$ wins election
$\parpair{C,V}$
  if and only if $\|V\| \geq 2$, $c$ is ranked first by the first voter in $V$
(recall that $V$ is input as a list, and so it is meaningful to speak
 of the first voter), and  ($c$ is ranked
  first by the second voter
  in $V$ or $c$ is ranked first or second by
  all voters).
It is immediate that $\election_1$ is candidate-anonymous
 and susceptible to
  constructive control by deleting candidates.
The following polynomial-time algorithm
shows that $\election_1$ is vulnerable to constructive control by
deleting candidates:
\begin{algorithmus}
\item Given an instance $\parpair{C,c,V,{k}}$ of Constructive Control by Deleting
  Candidates, if $\|V\| < 2$, then reject and halt. Otherwise,
  let $C_1$ be the set of all candidates preferred to $c$ by
  the first voter of $V$ and let $C_2$ be the set of all candidates
  preferred to $c$ by the second voter of $V$.
\item If $\|C_1 \cup C_2\| \leq {k}$ then accept and halt
  ($c$ can be made the unique winner by deleting $C_1 \cup C_2$).
\item If $\|C_1 \cup C_2\| > {k}+1$ then reject and halt
  (to ensure that $c$ is the unique winner, it is necessary (but not sufficient)
   to delete all candidates in $C_1$ and all but one of the candidates in
$C_2$).
\item \label{alg:ccdc-in-p} If $\|C_1 \cup C_2\| = {k}+1$ then for each $d
  \in C_2$, consider the election with candidates $(C - (C_1 \cup C_2)) \cup
  \{d\}$ and voters $V$.  If
  $c$ wins this election (i.e., is in first or second position for each
  voter), then accept and halt.
\item If none of the iterations of the for-each loop in
  Step~\ref{alg:ccdc-in-p} was successful, then reject and halt.
\end{algorithmus}

It remains to prove that $\hybrid(\election_0,\election_1)$ is resistant
to constructive control by deleting candidates.
Susceptibility follows from Theorem~\ref{t:ABK}.
To show NP-hardness,
we provide a $\manyonetext$-reduction from the
standard $\np$-complete problem Vertex Cover (see Garey and
Johnson~\cite{gar-joh:b:int}), which is defined as follows:

\subsubsection*{Vertex Cover}

\begin{description}
\item[Given:] An undirected, simple graph $G$ and a positive
  integer~$k$.
\item[Question:] Does $G$ have a vertex cover of size at most~$k$, i.e., does
  there exist a subset $V'$ of $G$'s vertex set such that $\|V'\| \leq k$ and
  $V'$ contains at least one vertex of each edge in~$G$?
\end{description}

Given an instance $(G,k)$ of Vertex Cover, create an instance 
$\parpair{C,0,V,k}$ 
of Constructive
Control by Deleting Candidates for election system
$\hybrid(\election_0,\election_1)$ as follows.  Let $G$ have $n$ vertices
and $m$ edges.
Without loss of generality, let the vertices of $G$ be $2,4,\ldots, 2n$.
The candidate set is defined by $C = \{0,1,2,4,6, \ldots , 2n\}$, where
$0$ is the distinguished candidate, $1$ is the candidate
whose presence determines whether $\hybrid$ routes its input to
$\election_0$ or $\election_1$, and 
$2, 4, 6, \ldots , 2n$ are candidates corresponding to the vertices of~$G$.
The voters in $V$ are defined as follows:
\begin{enumerate}
\item The first voter has preference $0 > \cdots$,
\item the second voter has preference $1 > 0 > \cdots$, and
\item there are $m$ additional voters, corresponding to the edges
of $G$:
For every edge $\{i,j\}$ in $G$ with $i < j$, we add a voter with
preference $i > j > 0 >  \cdots$.
\end{enumerate}
In all the above preferences, the remaining candidates follow in some
arbitrary order.  Clearly, this transformation is polynomial-time computable.
To show that Vertex Cover $\manyonetext$-reduces to Constructive Control by
Deleting Candidates for $\hybrid(\election_0,\election_1)$,
we prove the following claim.

\begin{claim}
\label{cla:t:one-off}
$G$ has a vertex cover of size $\leq k$ ($k \leq n$)
if and only if candidate~$0$ can be made
the unique winner of election $\parpair{C,V}$ with respect to system
$\hybrid(\election_0,\election_1)$ by deleting at most $k$ candidates.
\end{claim}

\sproofof{Claim~\ref{cla:t:one-off}}
From left to right: Delete the candidates corresponding to a vertex cover
of~$G$.  Since candidate~$1$ has not been deleted,
$\hybrid(\election_0,\election_1)$  routes its input to the system
$\electiondefault = \election_1$. Thus, candidate~$0$ is the unique winner.

From right to left: If candidate~$0$ is the unique winner after at most $k$
candidates have been deleted, candidate~$1$ cannot be among the deleted
candidates. (Note that otherwise $\hybrid(\election_0,\election_1)$ would have
routed its input to the system~$\election_0$, and since after deleting at most
$k \leq n$ candidates at least two candidates remain, no one would have won
in~$\election_0$.)  Since candidate 1 is ranked first by the second voter,
it follows that candidate~$0$ must be ranked first or
second by all voters.  This means that the set of deleted candidates
corresponds to a vertex cover of $G$ having size at 
most~$k$.~\eproofof{Claim~\ref{cla:t:one-off}}

This completes the proof.~\eproofof{Theorem~\ref{t:one-off}}

\sproofof{Theorem~\ref{t:let}}
The proof is similar to the proof of Theorem~\ref{t:nonmess}.
In all eight cases, it follows immediately from Theorem~\ref{t:ABK}
that $\hybrid(\election_0, \election_1, \ldots , \election_{k-1})$
is susceptible to $\Phi$. 
Since $\p = \np$ and the winner problem in each of 
Case~\ref{t:let:was-part-a-1} through~\ref{t:let:was-part-a-4}
and the unique winner problem in each of 
Case~\ref{t:let:was-part-b-1} through~\ref{t:let:was-part-b-4}
for $\hybrid(\election_0, \election_1, \ldots , \election_{k-1})$ is
in the polynomial hierarchy, we have a polynomial-time
algorithm for $\hybrid(\election_0, \election_1, \ldots , \election_{k-1})$'s
winner problem in each of 
Case~\ref{t:let:was-part-a-1} through~\ref{t:let:was-part-a-4} and 
a polynomial-time
algorithm for $\hybrid(\election_0, \election_1, \ldots , \election_{k-1})$'s
unique winner problem in each of 
Case~\ref{t:let:was-part-b-1} through~\ref{t:let:was-part-b-4}.

To prove vulnerability to the given control type~$\Phi$,
nondeterministically guess every possible way of 
partitioning the candidate set and
use the polynomial-time algorithm for
$\hybrid(\election_0, \election_1, \ldots , \election_{k-1})$'s winner
problem if ties in subelections are handled by~TP and
use the polynomial-time algorithm for
$\hybrid(\election_0, \election_1, \ldots , \election_{k-1})$'s unique winner
problem if ties in subelections are handled by~TE to determine
which candidates participate in the final round of the election.
Then use a polynomial-time algorithm for the unique winner problem
(note that when a winner problem is in~$\p$, that
clearly implies that the corresponding unique winner problem is in~$\p$)
to determine if the distinguished candidate is the unique
winner of the (final round of the) election for constructive control cases, and 
if the distinguished candidate is not a unique
winner of the (final round of the) election for destructive control
cases.~\eproofof{Theorem~\ref{t:let}}

\sproofof{Theorem~\ref{t:edith-147.A}}
  Let $\election_0$ be the election system in which each candidate $c$ is a
  winner if and only if $c$ is the only candidate and there is only one voter.
  This artificial election system is clearly vulnerable
  to constructive control by run-off partition of candidates.  To define
  $\election_1$, we modify election system $\election_1$
  from the proof of Theorem~\ref{t:one-off}.
  Here, we define $\election_1$ to be the following artificial
  election system: 
  For each candidate $c \in C$, $c$ wins election
$\parpair{C,V}$ if and only if
\begin{enumerate}
\item for some integer $n \geq 0$,
 $\|V\| = 1 + n(n-1)/2$ and $\|C\| \leq
  (n+3)/2$, and
\item $c$ is ranked first by the first voter in $V$ and 
  $c$ is ranked first or second by all voters.
\end{enumerate}
It is immediate that $\election_0$ and $\election_1$ are candidate-anonymous.
Note that tie-handling issues do not apply, since
$\election_0$ and $\election_1$ never have ties.
It is easy to see that $\election_1$ is susceptible to constructive control by
run-off partition of candidates.
We now describe a polynomial-time algorithm to decide if this type of control
can be asserted.

Given an instance $\parpair{C,c,V}$ of Constructive Control by
Run-off Partition of Candidates,
first check that for some integer $n \geq 0$, $\|V\| = 1 + n(n-1)/2$.
If not, for no partition of candidates is $c$ the unique winner.
  
If $\|V\| = 1 + n(n-1)/2$ for some integer $n \geq 0$, we claim that
$c$ is the unique winner in partition $\parpair{\{c\},C - \{c\}}$ if
and only if
there is a candidate partition  $\parpair{C_1,C_2}$ in which
$c$ is the unique winner.
This proves that $\election_1$'s
control problem is in P, since checking if $c$ is the unique winner
in $\parpair{\{c\},C - \{c\}}$ can be done in polynomial time.  

To prove the nontrivial direction of our
claim, suppose that 
$c$ is the unique winner in  partition  $\parpair{C_1,C_2}$
where $c \in C_1$, but
that $c$ is not the unique winner in partition $\parpair{\{c\},C - \{c\}}$. 
Since $c$ is the unique winner of $\parpair{\{c\},V}$, there is a candidate
$d \in C - \{c\}$ such that $d$ is the unique winner of
$\parpair{C-\{c\},V}$ and such that $c$ is not the unique winner of
$\parpair{\{c,d\},V}$. It follows that $d$ is ranked first by the first voter.
If $d \in C_1$, then $c$ is not the unique winner of $\parpair{C_1,V}$, which
contradicts the assumption that $c$ is the unique winner in
partition $\parpair{C_1,C_2}$.  If $d \in C_2$, then $d$ is the unique winner
of $\parpair{C_2,V}$ (since $C_2 \subseteq C - \{c\}$ and $d$ is the
unique winner of $\parpair{C-\{c\},V}$). It follows that
the final run-off is $\parpair{\{c,d\}, V}$.  Since $d$ is ranked first
by the first voter, $c$ does not win the final run-off, 
contradicting the assumption that $c$ is the unique winner in
partition $\parpair{C_1,C_2}$.

\medskip

It remains to prove that
$\hybrid(\election_0,\election_1)$ is resistant to constructive control by
run-off partition of candidates. 
Susceptibility follows from Theorem~\ref{t:ABK}.
To show $\np$-hardness of the control problem, we provide a reduction from the
problem Odd Half Vertex Cover, which is a minor modification of Vertex Cover
and is defined as follows:

\subsubsection*{Odd Half Vertex Cover}
\begin{description}
\item[Given:] An undirected, simple graph $G$ with an odd number $n > 1$ of
  vertices.
\item[Question:] Does $G$ have a vertex cover of size 
$(n+1)/2$?
\end{description}
Standard padding can be used to show that Odd Half Vertex Cover is 
$\np$-hard (see, for example, \cite[Problem 7.22]{sip:b:introduction-second-edition}).
The reduction from Odd Half Vertex Cover to Constructive
Control by Run-off Partition of Candidates
is similar to the reduction from Vertex Cover to Constructive
Control by Deleting Candidates,
which was presented in the proof of Theorem~\ref{t:one-off}.  

Given an
instance $G$ of Odd Half Vertex Cover,
create an instance $\parpair{C,0,V}$ of Constructive
Control by Run-off Partition of Candidates
for election system
$\hybrid(\election_0,\election_1)$ as follows.
Let $G$ have $n > 1$ vertices, $n$ odd.
Without loss of generality, let the vertices of $G$ be $2,4,\ldots, 2n$.
The
candidate set is defined by $C = \{0,1,2,3,4,6,8, \ldots , 2n\}$, where $0$ is
the distinguished candidate, $1$ and $3$ are
candidates whose presence determines whether $\hybrid$ routes its input to
$\election_0$ or $\election_1$, and
$2, 4, 6, \ldots , 2n$ are candidates corresponding to the vertices of~$G$.
The $1 + n(n-1)/2$ voters in $V$ are defined as follows:
\begin{enumerate}
\item The first voter has preference $0 > 3 > \cdots $,
\item for every edge $\{i,j\}$ in $G$ with $i < j$, we add
a voter with preference $3 > i > j > 0 > \cdots$,
and
\item add duplicates of the first voter until there are exactly $1 + n(n-1)/2$
voters.
\end{enumerate}
In the above preferences, the remaining candidates follow in some
arbitrary order. 
Clearly, this transformation is polynomial-time computable.
To show that Odd Half Vertex Cover $\manyonetext$-reduces to
Constructive Control by Run-off Partition of Candidates 
for $\hybrid(\election_0,\election_1)$, we prove the following claim.

\begin{claim}
\label{cla:edith-147.A}
$G$ has a vertex cover of size $(n+1)/2$
if and only if candidate~$0$ can be made
the unique winner of election $\parpair{C,V}$ with respect to system
$\hybrid(\election_0,\election_1)$ by run-off partition of candidates.
\end{claim}

\sproofof{Claim~\ref{cla:edith-147.A}}
From left to right:   Let $C' \subseteq \{2,4,6,\ldots, 2n\}$ be a vertex
cover of $G$ of size $(n+1)/2$.  Consider candidate partition
$C_2 = \{3\} \cup C'$ and $C_1 = C - C_2$.  Note that
$\|C_1\| = \|C_2\| = (n+3)/2$.  Also note that both subelections
are routed to $\election_1$.  It is immediate that
$3$ is the unique winner of $\parpair{C_2,V}$.
Since $C'$ is a vertex cover of $G$, $0$ is the unique
winner of $\parpair{C_1,V}$. It is immediate
that $0$ is the unique winner of the final run-off 
$\parpair{\{0,3\},V}$.

From right to left: Suppose $0$ is the unique winner in partition
$\parpair{C_1,C_2}$ where $0 \in C_1$. 
Then subelection $\parpair{C_1,V}$ and the final run-off need to involve
candidates from $\{1,3\}$ (since $\|V\| \neq 1$, subelections routed to
$\election_0$ will not have winners).  
It follows that $0$ is the unique winner of $\parpair{C_1,V}$, that
$1$ or $3$ is the unique winner of $\parpair{C_2,V}$, and that
exactly one of $1$ or $3$ is in $C_1$.  Since both subelections
have unique winners, it follows that
$\|C_1\| = \|C_2\| = (n+3)/2$.  Since $0$ is the unique winner
of $\parpair{C_1,V}$, the vertex candidates that are not in $C_1$
(i.e., that are in $C_2$) form a vertex cover of $G$.
$C_2$ contains exactly
$(n+3)/2 - 1 = (n+1)/2$  vertex candidates.
So, $G$ has a vertex cover of size 
$(n+1)/2$.~\eproofof{Claim~\ref{cla:edith-147.A}}

This completes the proof.~\eproofof{Theorem~\ref{t:edith-147.A}}

\sproofof{Theorem~\ref{t:partition}}   
As in the proof of Theorem~\ref{t:edith-147.A},
  let $\election_0$ be the election system in which each candidate $c$ is a
  winner if and only if $c$ is the only candidate and there is only one voter.
  This artificial election system is clearly vulnerable
  to constructive control by  partition of candidates. 
  To define $\election_1$, we modify election system $\election_1$
  from the proof of Theorem~\ref{t:edith-147.A}.
  Here, we define $\election_1$ to be the following artificial
  election system: 
  For each candidate $c \in C$, $c$ wins election
$\parpair{C,V}$ if and only if
\begin{enumerate}
\item for some even integer $n \geq 0$,
 $\|V\| = 1 + n(n-1)/2$ and $\|C\| \in \{1, n/2 + 2\}$, and
\item $c$ is ranked first by the first voter in $V$ and 
  $c$ is ranked first or second by all voters.
\end{enumerate}
It is immediate that $\election_0$ and $\election_1$ are candidate-anonymous.
Note that tie-handling issues do not apply, since
$\election_0$ and $\election_1$ never have ties.
It is easy to see that $\election_1$ is susceptible to constructive control by
partition of candidates.
We now describe a polynomial-time algorithm to decide if this type of control
can be asserted for~$\election_1$.

Given an instance $\parpair{C,c,V}$ of Constructive Control by
Partition of Candidates,
first check that for some even integer $n \geq 0$, $\|V\| = 1 + n(n-1)/2$.
If not, for no partition of candidates is $c$ the unique winner.

If $\|V\| = 1 + n(n-1)/2$ for some even integer $n \geq 0$,  
first check if $c$ is the unique winner in partition
$\parpair{C-\{c\},\{c\}}$.
Now suppose that
$c$ is not the unique winner in this partition.
Since $c$ is the unique winner of $\parpair{\{c\},V}$,
it follows that $\parpair{C-\{c\},V}$ has a winner,
which implies that  $\|C- \{c\}\| \in \{1, n/2 + 2\}$, i.e.,
$\|C\| \in \{2, n/2 + 3\}$.  

Note that if $c$ is the unique winner
in partition $\parpair{C_1,C_2}$, one of the following cases
holds: 
\begin{enumerate}
\item
$c \in C_1$ and $\|C_1\| \in \{1, n/2 + 2\}$,
\item
$c \in C_2$ and $\|C_2\| \in \{1, n/2 + 2\}$, or
\item
$c \in C_2$ and $\|C_2\| = n/2 + 1$.
\end{enumerate}
Since $\|C\| \in \{2, n/2 + 3\}$, 
we can enumerate all partitions $\parpair{C_1,C_2}$ that satisfy
one of these cases in polynomial time.  Control can be asserted
if and only if $c$ is the unique winner for one of these partitions.

\medskip

It remains to prove that
$\hybrid(\election_0,\election_1)$ is resistant to constructive control by
partition of candidates. 
Susceptibility follows from Theorem~\ref{t:ABK}.
To show $\np$-hardness of the control problem, we provide a reduction from the
NP-complete problem Even Half Vertex Cover, a minor variation of
the problem Odd Half Vertex Cover that was used in the proof of
Theorem~\ref{t:edith-147.A}.

\subsubsection*{Even Half Vertex Cover}
\begin{description}
\item[Given:] An undirected, simple graph $G$ with an even number $n > 0$ of
  vertices.
\item[Question:] Does $G$ have a vertex cover of size 
$n/2$?
\end{description}
The reduction from Even Half Vertex Cover to Constructive
Control by Partition of Candidates
is similar to the reduction from Odd Half Vertex Cover to Constructive
Control by Run-off Partition Candidates,
which was presented in the proof of Theorem~\ref{t:edith-147.A}.

Given an
instance $G$ of Even Half Vertex Cover,
create an instance $\parpair{C,0,V}$ of Constructive
Control by Partition of Candidates
for election system
$\hybrid(\election_0,\election_1)$ as follows.
Let $G$ have $n > 0$ vertices, $n$ even.
Without loss of generality, let the vertices of $G$ be $2,4,\ldots, 2n$.
The
candidate set is defined by $C = \{0,1,2,4,6,8, \ldots , 2n\}$, where $0$ is
the distinguished candidate, $1$ is the
candidate whose presence determines whether $\hybrid$ routes its input to
$\election_0$ or $\election_1$, and
$2, 4, 6, \ldots , 2n$ are candidates corresponding to the vertices of~$G$.
The $1 + n(n-1)/2$ voters in $V$ are defined as follows:
\begin{enumerate}
\item The first voter has preference $0 > 1 > \cdots $,
\item for every edge $\{i,j\}$ in $G$ with $i < j$, we add
a voter with preference $i > j > 0 > \cdots$, and
\item add duplicates of the first voter until there are exactly $1 +
n(n-1)/2$ voters.
\end{enumerate}
In the above preferences, the remaining candidates follow in some
arbitrary order. 
Clearly, this transformation is polynomial-time computable.
To show that Even Half Vertex Cover $\manyonetext$-reduces to
Constructive Control by Partition of Candidates 
for $\hybrid(\election_0,\election_1)$, we prove the following claim.

\begin{claim}
\label{cla:partition}
$G$ has a vertex cover of size $n/2$
if and only if candidate~$0$ can be made
the unique winner of election $\parpair{C,V}$ with respect to system
$\hybrid(\election_0,\election_1)$ by partition of candidates.
\end{claim}

\sproofof{Claim~\ref{cla:partition}}
From left to right:   Let $C' \subseteq \{2,4,6,\ldots, 2n\}$ be a vertex
cover of $G$ of size $n/2$.  Consider candidate partition
$C_1 = C'$ and $C_2 = C - C_1$.  Note that $(C_1,V)$ has no winners, since
$\hybrid$ routes this election to $\election_0$.
Since $C'$ is a vertex cover of $G$ and $\|C_2\| = n/2 + 2$,
$0$ is the unique
winner of $\parpair{C_2,V}$, and thus of the whole election.

From right to left: Suppose $0$ is the unique winner in partition
$\parpair{C_1,C_2}$.  Then certainly $0$ is the unique winner
of some subelection.  Note that $n>0$ and $n$ is even implies $\|V\| > 1$.
So, in order for $0$ to win a subelection,
the subelection must be routed to $\election_1$,
which implies that 1 participates in the subelection. But this
implies that the number of candidates involved in the
subelection is $n/2 + 2$ and $0$ is ranked first or second
by all voters.  
It follows that the set of candidates not
involved in the subelection form a vertex cover of $G$,
and this vertex cover is of size $n/2$.~\eproofof{Claim~\ref{cla:partition}}

This completes the proof.~\eproofof{Theorem~\ref{t:partition}}

\sproofof{Theorem~\ref{t:destructive-vulnerability-not-inherits}}
In $\election_0$, each candidate $c$ wins if and only if there
is at least one voter and  $c$ is ranked
first by the first voter and (there are at least two voters or
there are at least two candidates).
This artificial system is clearly candidate-anonymous and
it is easy to see that $\election_0$ is
susceptible to destructive control by deleting candidates, 
partition of candidates (in models TE and~TP), and
run-off partition of candidates (in models TE and~TP): Consider
an election with two candidates, $c$ and $d$, and one voter, $c > d$.
Then $c$ is the unique winner of the election, but deleting $d$ or
partitioning $\{c,d\}$ into $\{c\}$ and $\{d\}$ will ensure that
$c$ does not win.  To show vulnerability, note that distinguished
candidate $c$ can be made not a unique winner of 
$\parpair{C,V}$ if and only if $\|V\| = 0$ or
$c$ is not ranked first by the first
voter or $\|V\| = 1$ (in the latter
case by deleting all candidates
other than $c$, or by partitioning the set of candidates $C$ into
$C_1 = \{c\}$ and $C_2 = C - \{c\}$).

In $\election_1$, for each candidate $c \in C$,
$c$ wins election $\parpair{C,V}$ if and only if
$\|V\| > 0$ and
$c$ is ranked second by the first voter and
($\|V\| \neq 4 \|C\|^2$ or some voter ranks $c$  worse than second).

It is immediate that $\election_1$ is candidate-anonymous.
Note that tie-handling issues do not apply, since
$\election_0$ and $\election_1$ never have ties.
It is easy to see that $\election_1$ is susceptible to destructive
control by deleting candidates, partition of candidates, and
run-off partition of candidates. To show vulnerability,
it suffices to show that
we can decide in polynomial time whether these types of control
can be asserted.

Given an instance $\parpair{C,c,V,{k}}$ of Destructive
Control by Deleting Candidates, control can
be asserted if
and only if
\begin{enumerate}
\item $\|V\| = 0$, or
\item $c$ is not ranked second by the first voter, or
\item $\|V\| = 4\|C\|^2$ and no voter ranks $c$ worse than second, or
\item $k > 0$.
\end{enumerate}
Note that if any of the first three cases holds, $c$ does not win, 
so no action by the chair is required to assert control. 
On the other hand, if none of these first three cases holds, 
the chair can block $c$ from winning exactly if $k>0$,
by deleting one candidate to make sure that 
$c$ is not ranked second by the first voter.

Given an instance $\parpair{C,c,V}$ of Destructive
Control by Partition of Candidates or of Destructive
Control by Run-off Partition of Candidates, note that
$c$ is not a winner in election $\parpair{\{c\},V}$.
Thus, these two types of
destructive control can always be asserted, using
partition $\parpair{\{c\},C-\{c\}}$.

\medskip

It remains to prove that
$\hybrid(\election_0,\election_1)$ is resistant to destructive control by
by deleting candidates, partition of candidates, and
run-off partition of candidates. 

Susceptibility follows from Theorem~\ref{t:ABK}.
To show $\np$-hardness of the control problem, we provide a reduction from the
NP-complete problem Even Half Vertex Cover, which was defined in the
proof of Theorem~\ref{t:partition}. 
The reduction is similar to the reduction from
Even Half Vertex Cover to Constructive Control by Partition of
Candidates from 
the proof of Theorem~\ref{t:partition}. 

Given an instance $G$ of Even Half Vertex Cover,
create an instance of the relevant destructive control problem
for election system $\hybrid(\election_0,\election_1)$ as follows.
Let $G$ have $n > 0$
vertices, $n$ even.
Without loss of generality, let the vertices of $G$ be $\{2,4,\ldots, 2n\}$.
The
candidate set is defined by $C = \{0,1,2,4,6,8, \ldots , 2n\}$, where $0$ is
the distinguished candidate, $1$ is the
candidate whose presence determines whether $\hybrid$ routes its input to
$\election_0$ or $\election_1$, and
$2, 4, 6, \ldots , 2n$ are candidates corresponding to the vertices of~$G$.
The $4(n/2+2)^2$ voters $V$ are defined as follows:
\begin{enumerate}
\item The first voter has preference $1 > 0 > \cdots $,
\item for every edge $\{i,j\}$ in $G$ with $i < j$, we add
a voter with preference $i > j > 0 > \cdots$, and
\item add duplicates of the first voter until there are exactly
$(n+4)^2 = 4(n/2 + 2)^2$ voters.
\end{enumerate}
In the above preferences, the remaining candidates follow in some
arbitrary order. 
Clearly, this transformation is polynomial-time computable.
To show that Even Half Vertex Cover $\manyonetext$-reduces to
Destructive Control by
Deleting Candidates,  Partition of Candidates, and
Run-off Partition of
Candidates for $\hybrid(\election_0,\election_1)$, it suffices
to prove the following claim.

\begin{claim}
\label{cla:destructive-vulnerability-not-inherits}
\begin{enumerate}
\item $G$ has a vertex cover of size $n/2$
if and only if candidate~$0$ can be made not a unique winner of
election $\parpair{C,V}$ with respect to system
$\hybrid(\election_0,\election_1)$ by deleting at most $n/2$ candidates.

\item $G$ has a vertex cover of size $n/2$
if and only if candidate~$0$ can be made
not a unique winner of election $\parpair{C,V}$ with respect to system
$\hybrid(\election_0,\election_1)$ by partition of candidates.

\item $G$ has a vertex cover of size $n/2$
if and only if candidate~$0$ can be made
not a unique winner of election $\parpair{C,V}$ with respect to system
$\hybrid(\election_0,\election_1)$ by run-off partition of candidates.
\end{enumerate}
\end{claim}

\sproofof{Claim~\ref{cla:destructive-vulnerability-not-inherits}}
From left to right (for all three cases):
Let $C' \subseteq \{2,4,6,\ldots, 2n\}$ be a vertex
cover of $G$ of size $n/2$.  Consider election $(C-C',V)$.
Since $\|V\| = 4(n/2 + 2)^2 = 4 \|C-C'\|^2$ and $0$
is ranked first or second
by all voters in $(C-C',V)$, it follows that $0$ is not a winner
of $(C-C',V)$.  Thus, destructive control can be asserted by deleting
the $n/2$ candidates in $C'$ in the deleting candidates case, and
by partitioning the candidates as $\parpair{C-C',C'}$ in the partition
and run-off partition of candidates cases.

From right to left (for all three cases): Suppose $0$ can be made
not a unique winner of election $\parpair{C,V}$
by deleting candidates, or by partition of candidates, or by run-off
partition of candidates.  Then it
certainly 
follows that for
some $C' \subseteq C$ with $0 \in C'$, $0$ is not a winner
of $\parpair{C',V}$. Note that candidate 1 must participate in 
election $\parpair{C',V}$, since $0$ is the unique winner of all
subelections that don't involve $1$: such elections are routed to
$\election_0$, $\|V\| \geq 2$, and the first voter ranks $0$ first
of the candidates in $C - \{1\}$.

So, $0$ is not a winner of $\parpair{C',V}$ for
some $C'$ such that $\{0,1\} \subseteq C' \subseteq C$.
Since $\parpair{C',V}$ is routed to $\election_1$ and
$0$ is ranked second by the first voter in $\parpair{C',V}$,
it follows that $\|V\| = 4\|C'\|^2$ and that $0$ is  
ranked first or second by all voters.  The latter condition
implies that the set of candidates not involved in the subelection, i.e.,
the candidates in $C - C'$,  form a vertex cover of $G$.  Since
$\|V\| = 4(n/2+2)^2$ and $\|V\| = 4\|C'\|^2$, it follows that
$\|C'\| = n/2 + 2$, and thus the vertex cover of $G$,
$C - C'$, has size 
$n/2$.~\eproofof{Claim~\ref{cla:destructive-vulnerability-not-inherits}}

This completes the 
proof.~\eproofof{Theorem~\ref{t:destructive-vulnerability-not-inherits}}

\end{document}